\pdfoutput=1
\documentclass[numberedappendix,revtex4,iop]{emulateapj}
\usepackage{hyperref}
\usepackage{amssymb,amsmath}
\usepackage{refcount}
\newif\ifAMStwofonts

\def\sqiglt{\hbox{\rlap{\lower.55ex \hbox {$\sim$}}\kern-.05em \raise.4ex \hbox{$<$}\,}}
\def\sqiggt{\hbox{\rlap{\lower.55ex \hbox {$\sim$}}\kern-.05em \raise.4ex \hbox{$>$}\,}}
\def\til{\ensuremath{\sim\,}}

\def\chisq{\ensuremath{\chi^2}}
\def\rchisq{\ensuremath{\chi_{\nu}^{2}}}
\newcommand{\tim}[1]{\ensuremath{\times 10^{#1}}}
\def\deg{\ensuremath{^{\circ}}}

\def\swift{\emph{Swift}}
\def\good{\emph{Good}}
\def\reasonable{\emph{Reasonable}}
\def\poor{\emph{Poor}}
\def\arcmin{\ensuremath{^{\prime}}}
\def\arcsec{\ensuremath{^{\prime\prime}}}
\def\cms{\ensuremath{$cm$^{-2}}}
\def\cstat{\ensuremath{\mathcal{C}}}

\begin{document}

\title{1SXPS: A deep Swift X-ray Telescope point source catalog with light curves and spectra}
\shorttitle{The 1SXPS catalog}

\shortauthors{Evans et al.}

\author{P.A. Evans, J.P. Osborne, A.P. Beardmore, K.L. Page, R. Willingale, C.J. Mountford, C. Pagani}
\affil{University of Leicester, X-ray and Observational Astronomy Group, Department of Physics and Astronomy,
University Road, Leicester, LE1 7RH, UK}
\email{pae9@leicester.ac.uk}


\author{D.N. Burrows and J.A. Kennea}
\affil{Department of Astronomy and Astrophysics, Pennsylvania State University, University Park, Pennsylvania 16802, USA}


\author{M. Perri}
\affil{ASI-Science Data Center, Via del Politecnico, I-00133 Rome, Italy 
\\
INAF-Osservatorio Astronomico di Roma, Via Frascati 33, I-000040 Monteporzio Catone, Italy}


\author {G. Tagliaferri}
\affil{INAF – Osservatorio Astronomico di Brera, via E. Bianchi 46, 23807 Merate (LC), Italy }

\and

\author {N. Gehrels}
\affil{NASA/Goddard Space Flight Center, Greenbelt, MD 20771, USA}

\label{firstpage}

\begin{abstract}  
We present the 1SXPS (Swift-XRT Point Source) catalog of 151,524 X-ray point-sources detected by the \swift-XRT in 8 years 
of operation. 
The catalog covers 1905 square degrees distributed approximately uniformly on the sky. We analyze the data in two ways. First we
consider all observations individually, for which we have a typical sensitivity of \til3\tim{-13} erg \cms\ s$^{-1}$ (0.3--10 keV).
Then we co-add all data covering the same location on the sky: these images have a typical sensitivity of \til9\tim{-14} erg \cms\ s$^{-1}$ (0.3--10 keV).
Our sky coverage is nearly 2.5 times that of 3XMM-DR4, although the catalog is a factor of \til1.5 less sensitive.
The median position error is 5.5\arcsec (90\% confidence), including systematics.
Our source detection method improves on that used in previous XRT catalogs and we report $>68,000$ new X-ray sources.
The goals and observing strategy of the \swift\ satellite allow us to probe source variability on multiple timescales,
and we find $\til30,000$ variable objects in our catalog. For every source we give positions, fluxes, time series
(in four energy bands and two hardness ratios), estimates of the spectral properties, spectra and spectral
fits for the brightest sources, and variability probabilities in multiple energy bands and timescales.
\end{abstract}

\keywords{Catalogs -- Surveys -- X-rays: general -- Methods: data
analysis}

\section{Introduction}
\label{sec:intro}

\begin{figure}
\begin{center}
\includegraphics[width=8.1cm]{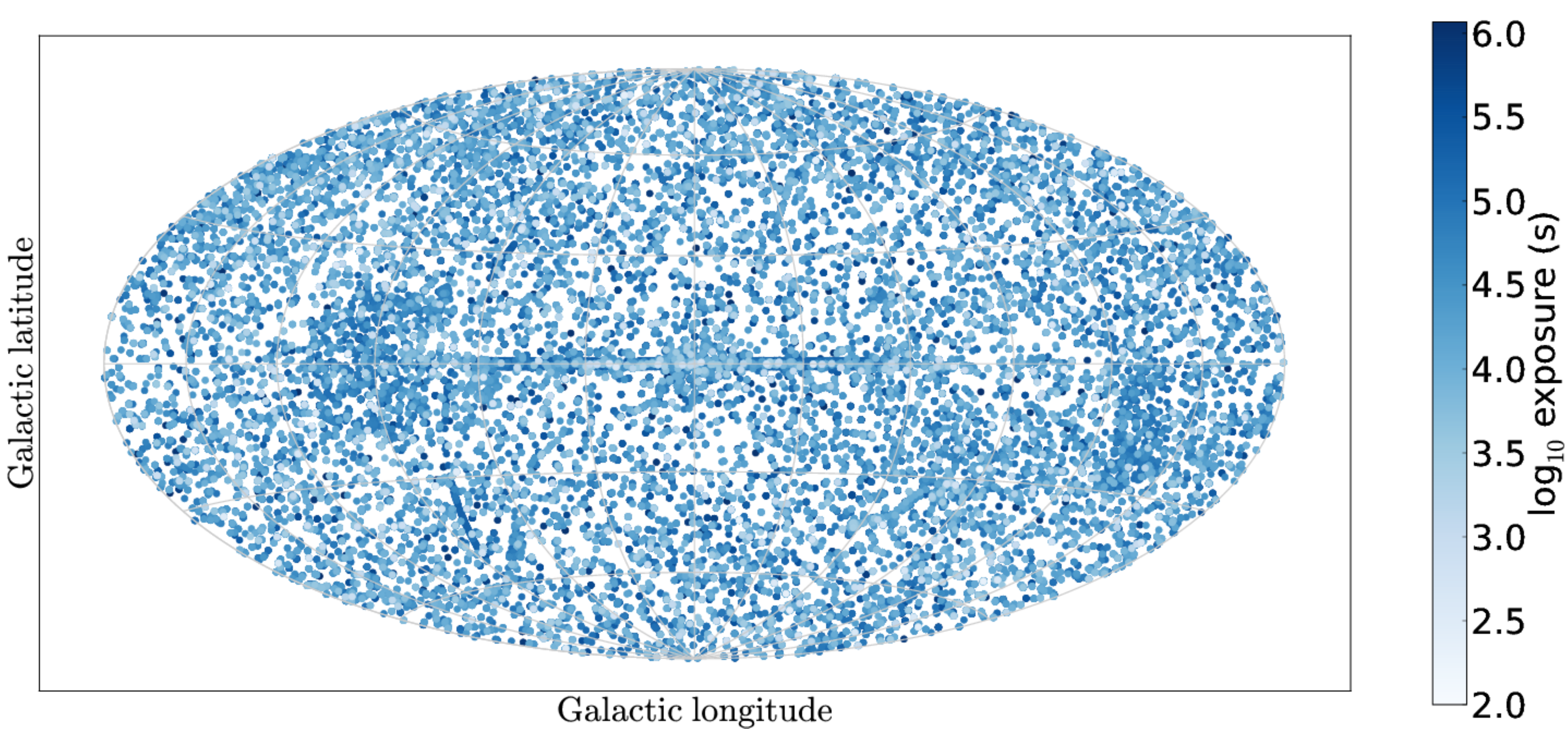}
\caption{The locations of the observations in the 1SXPS catalog in Galactic coordinates. The 
colors of the points indicate the exposure time included in the catalog. The point sizes are larger than the XRT field of view.
A color version of this figure is available in the electronic version of
the paper.}
\label{fig:skyplot}
\end{center}
\end{figure}

Serendipitous X-ray source catalogs have been produced for most X-ray
satellites since the \emph{Einstein\/} mission (e.g.\ \citealt{Gioia90,Voges99,Ueda05,Watson09,iEvans10})
and have contributed much to our understanding of the X-ray sky. 
The \swift\ satellite \citep{GehrelsSwift} has several unique features which
mean that a serendipitous source catalog produced from its X-ray
Telescope (XRT; \citealt{BurrowsXRT}) can make a distinctive
contribution to this field, particularly in the area of source variability.
To make this catalog we have analyzed \swift-XRT data from the first 8 years of operations, 
covering 13,065 distinct locations (giving a coverage of 1905 square degrees),
of which 81\%\ were observed at least twice. In many cases a field is observed
both multiple times within a day, and over a period of many days, allowing us to probe
variability on different timescales. Swift pointings have been performed across the entire sky with considerable 
uniformity, although there is an over-density of pointings along the 
Galactic plane; see Fig.~\ref{fig:skyplot}.

The XRT contains a CCD detector with a bandpass of 0.3--10 keV, with a peak effective area of 110 cm$^2$ at 1.5 keV.
The field of view has a radius of 12.3\arcmin, with vignetting at the outer edge reducing the effective
area by \til 25\% (at 1.5 keV); there are also several detector columns permanently masked out due to damage
from a micrometeoroid impact of 2005 May 27\citep{Abbey06}.

Two previous XRT point-source catalogs have been produced, which used the routines built into
the {\sc ximage} software to detect sources. The first, \cite{puccetti11}, analyzed the deepest GRB
fields, combining all of the data into a single image per field. The second, \cite{delia13}, analyzed 7 years of XRT 
data, considering each observation independently. For this catalog we have developed a new detection method
capable of detecting fainter sources than 
these papers, and have conducted a rigorous analysis of our completeness
and false positive rate; we have also considered both individual observations and deep images,
making this a more complete point source catalog than those of \cite{puccetti11} and \cite{delia13}. 
We have produced light curves and variability estimates for every source detected in the catalog.
These are available through a dedicated website.

We performed our analysis in four energy bands: one covering the entire 
calibrated energy range of the XRT (0.3--10 keV), and three partial bands which were chosen to overlap those used in
the 2XMM catalog \citep{Watson09}; these are listed in Table~\ref{tab:summary}.
For a typical AGN spectrum this will give approximately the same number of events in each of the
three partial bands. Summary details of the catalog are given in Table~\ref{tab:summary}.

This paper is organized as follows: In Section~2 we discuss the data selection and filtering
applied before collating the catalog. In Section~3 we detail the analysis process, the results of which are given in 
Section~4. In Section~5 we demonstrate the reliability of our catalog compilation, while Section~6 discusses
the false positive rate and completeness.

\begin{deluxetable*}{ccc}
\tablecaption{Summary details of the catalog}
 \tablehead{
 \colhead{Category}      & \colhead{Units}   &    \colhead{Value} 
}
\startdata
Energy Bands:  & keV & Total =  $0.3\le E \le 10$ \\
& &  Soft = $0.3\le E < 1$ \\
& &  Medium = $1\le E < 2$  \\
& &  Hard $2\le E < 10$ \\
Sky Coverage & square degrees & 1905  \\
Median sensitivity (0.3--10 keV) &  erg \cms\ s$^{-1}$ & 3\tim{-13}  \\
Number of detections & & 585,443 \\
Number of unique sources & &  151,524 \\
Number of variable sources & &  28,906 \\
Number of uncataloged sources$^1$ & & 68,638 \\
\enddata
\tablecomments{$^1$i.e. without a match within 3-$\sigma$ in any of the catalogs detailed
in Section~\ref{sec:xcorr} excluding the 2MASS and USNO-B1 catalogs.}
\label{tab:summary}
\end{deluxetable*}

\begin{figure}
\begin{center}
\includegraphics[width=8.1cm]{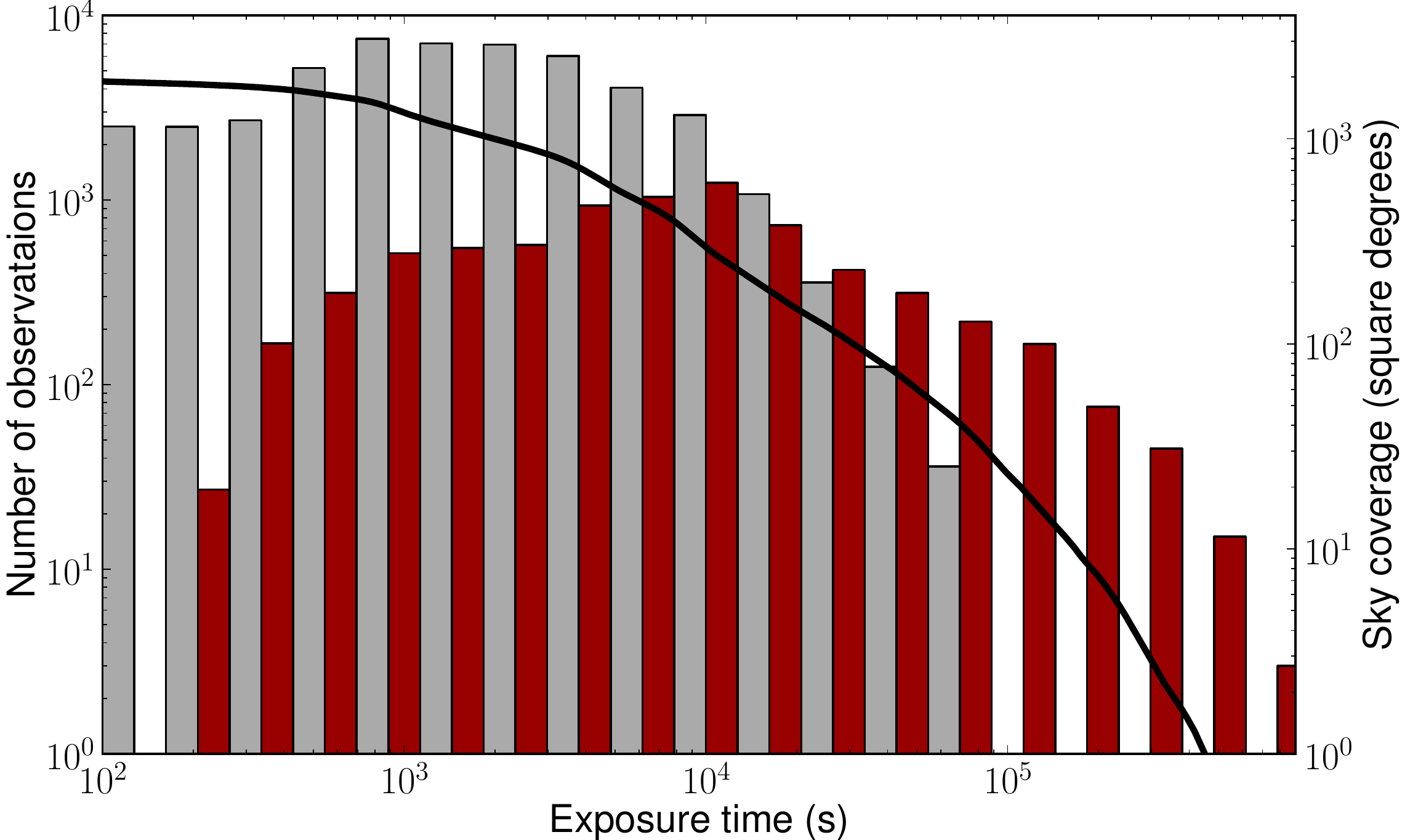}
\caption{Temporal and geometric coverage of the 1SXPS catalog. The solid line shows the 
unique sky coverage of the catalog as a function of exposure time. The histogram shows
the distribution of exposure times of the observations (gray) and the
stacked images (black; red in the electronic version).}
\label{fig:expo}
\end{center}
\end{figure}

\subsection{Data timescales: snapshots, observations and stacked images}
\label{sec:timescale}

\swift\ data are organized into \emph{snapshots} and \emph{observations}. Due to its low Earth orbit (P=96 min), 
\swift\ cannot observe an object continuously for more than 2.7 ks, 
thus most observations are spread over multiple spacecraft orbits. 
A single, continuous on-target exposure is referred to as a \emph{snapshot}. 
Within a UT day\footnote{i.e. from 00:00:00 to 23:59:59 UT.}, the data from all 
snapshots pointed at a given source are aggregated into a single dataset, 
referred to as an \emph{observation} and is assigned a unique 
\emph{ObsID\/} under which the data can be accessed. In order to probe 
source variability we consider both of these timescales. Neither snapshots
nor observations have a standard duration: snapshots may be 300--2700 s in duration\footnote{Shorter
snapshots are possible if a Gamma Ray Burst interrupts the planned observations.} and there are typically
1--15 snapshots in an observation. However snapshot-to-snapshot variability probes 
timescales $<$1 day, while observation to observation variability probes timescales $>$1 day.

Snapshots are generally too short for any but the brightest sources to be detected, therefore we search for sources
in each observation and on summed images comprising all XRT observations on each location 
of the sky. We refer to these latter as \emph{stacked images}. The word \emph{image\/} where it appears 
in this paper can be taken literally as a single (FITS) image, which may be of a snapshot, observation 
or a stacked image; whereas \emph{field\/} refers to an area on the sky. Fig.~\ref{fig:expo} shows the distribution
of exposure times in the two types of image on which we perform source detection,
and the sky coverage of the catalog as a function of exposure time.

\section{Data selection}
\label{sec:selection}

Initially we selected every XRT science observation\footnote{Excluding
ObsIDs beginning with `006', as these are calibration datasets, sometimes
taken in non-standard operating modes.} 
collected before 2012 October 12 containing at least 100 s of Photon 
Counting (PC) mode data\footnote{Windowed Timing mode data have only 1-D 
spatial resolution so are inappropriate for detecting and localizing sources.}; we also 
required that at least one snapshot in the observation was at least 100 s in 
duration. We removed any observations which overlap 
the locations listed in Table~\ref{tab:ignore}, as these include large-scale diffuse emission
(identified by examining the XRT images) which is not well handled by our 
point-source-optimized detection system. We then filtered the remaining event lists to remove time intervals
where the data were affected by
light reflected off the sunlit Earth, or where the astrometry was unreliable (both described below); if this reduced the exposure time to below the 100-s limit,
the observation was discarded.

\begin{deluxetable}{ccc}
\tablecaption{Locations excluded from the catalog due to large-scale emission structures}
 \tablehead{
 \colhead{RA}      &       \colhead{Dec}  & \colhead{Identity} \\
 \colhead{deg, J2000} & \colhead{deg, J2000} \\
}
\startdata
6.334 & 64.136 & Tycho SNR   \\
16.006 & -72.032 &  SNR B0102-72.3   \\
28.197 & 36.153 & RSCG15     \\
44.737 & 13.582 & ACO 401   \\
49.951 & 41.512 & NGC 1275 \\
81.510 & 42.942 & Swift J0525.8+4256 \\
83.633 & 22.014 & Crab Nebula   \\
83.867 & -69.270 & SN 1987A   \\
85.052 & -69.331 & PSR 0540-69   \\
94.277 & 22.535 & OFGL J0617.4+2234 \\
116.882 & -19.303 & PKS 0745-191   \\
125.851 & -42.781 & Pup A   \\
139.527 & -12.100 & Hydra A   \\
161.017 & -59.746 & Carina Nebula   \\
177.801 & -62.626 & ESO 130-SNR001   \\
187.709 & 12.387 & M87   \\
194.939 & 27.943 & Coma Cluster   \\
207.218 & 26.590 &  Abell 1795   \\
227.734 & 5.744 & Abell 2029   \\
229.184 & 7.020 & Abell 2052   \\
234.798 & -62.467 & Swift J1539.2-6227 \\
239.429 & 35.507 & Abell 2141  \\
244.405 & -51.041 & SNR G332.4-00.4   \\
258.116 & -23.367 & Ophiuchi Cluster   \\
266.414 & -29.012 & Galactic Center   \\
299.868 & 40.734 & 3C405.0   \\
326.170 & 38.321 & Cyg X-2   \\
345.285 & 58.877 & 1E2259+586   \\
350.850 & 58.815 & Cas A   \\
\enddata
\tablecomments{Observations within 12.5\arcmin\ of these locations are excluded from our catalog.}
\label{tab:ignore}
\end{deluxetable}

\subsection{Bright Earth filtering}
\label{sec:brightearth}

When \swift\ points close to the Earth limb, at certain spacecraft roll angles
the background level in the XRT is increased by contamination from light scattered off the sunlit side of the Earth.
This is always most notable on the left-hand side of the detector. For each observation
we therefore examined the raw event list (before the {\sc xrtpipeline} script has been executed) and selected events in
a box $122\times350$ pixels in size, centered on the XRT detector pixel $(62,300)$ (i.e. the left hand side). Times where the event rate
in this box exceeds 40 event s$^{-1}$ were deemed to be affected by bright Earth, and were removed from 
the observation before further processing. For 90\%\ of the observations in our catalog, this removed less
than 10\%\ of the exposure time.

\subsection{Astrometry filtering}
\label{sec:astromfilt}

The standard astrometric calibration of XRT data is taken from the \swift\ star 
trackers, mounted on the XRT. This provides a solution which is accurate to 3.5\arcsec\ 90\% of the 
time \citep{Moretti07}. We identified and removed times where this 
astrometry was incorrect by more than 10\arcsec\ by using the 
UV/Optical telescope (UVOT) on \swift. For each UVOT image we corrected 
the astrometry by matching UVOT sources to the USNO-B1 catalog. We 
then determined the magnitude of this correction on the X-ray sources 
in the image and at four locations positioned symmetrically in the field 
at radii of 5.9\arcmin\ from the field center (i.e.\ mid-way to the edge of the 
field). This was done using the known translation from the UVOT detector to the XRT 
detector, as described in \cite{Goad07}. If any of these corrections were $>$10\arcsec\ we marked the times of that UVOT 
image as bad and excluded XRT data taken during those times from the 
analysis. This was implemented as a two-pass process, since it 
makes use of the XRT source list for a given observation, which was not produced until
the entire detection system had completed. We therefore ran the detect 
procedure on the per-observation timescale in full without this phase
before performing this astrometric check. Any observations identified by 
this process were then reanalyzed from scratch, with the times of poor astrometry removed.
The stacked images were only created and processed after this had been completed.

\begin{figure*}
\begin{center}
\includegraphics[width=17cm]{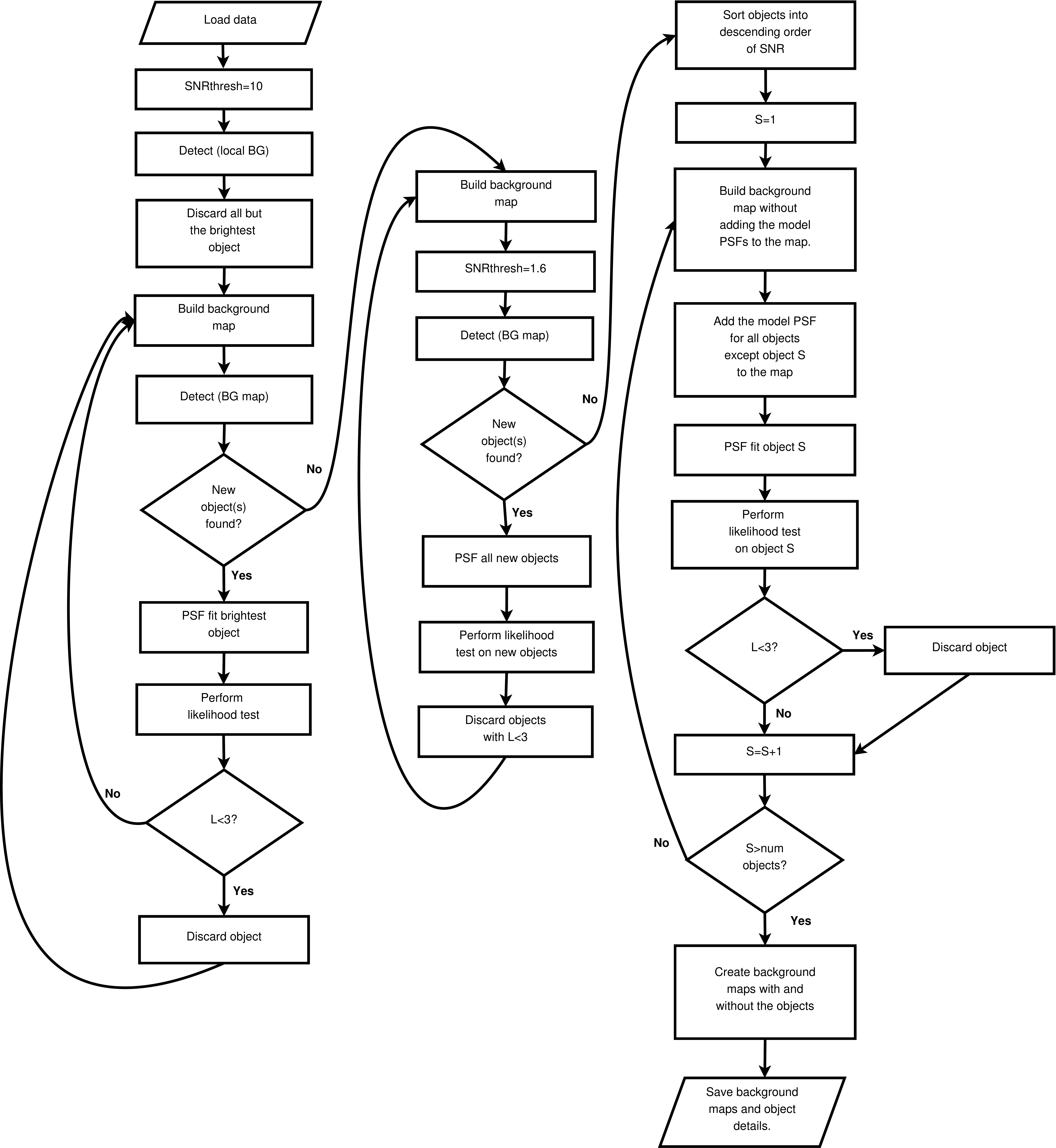}
\caption{Flowchart showing the source detection and characterization algorithm.}
\label{fig:procflowchart}
\end{center}
\end{figure*}

\section{Data Processing}
\label{sec:dataproc}

For all analysis in this catalog we used the {\sc heasoft} version 6.12 software which
includes the {\sc XRTDAS} v2.8.0 developed at the ASI Science Data Center (ASDC, Italy), and
the XRT CALDB version 20120209. Event files were reprocessed using the {\sc xrtpipeline} task with the standard filtering
criteria to provide a self-consistent and up-to-date set of event lists.

\subsection{Stacked image creation}
\label{sec:stackedim}

Our source detection software works in the {\sc sky} $(x,y)$ coordinate system, which is a virtual system
constructed using a tangent plane projection such that $(x,y)$ has a linear mapping to (RA, Dec) (see
\citealt{Greisen02,Calabretta02}). This coordinate system is produced uniquely for each ObsID when 
{\sc xrtpipeline} is run. For the stacked images we therefore used the {\sc coordinator ftool} to
reconstruct the coordinates for all observations within a stacked image using the same projection.


There is a small number of locations on the sky  (4\%\ of those covered
by our catalog) where overlapping observations exist that extend beyond the $1000\times1000$ pixel 
(=$39.9\arcmin\times39.3\arcmin$) range of the {\sc sky} coordinates in the XRT event files and thus could not be covered
by a single stacked image. In these cases we split the observations into multiple stacked images, aiming to minimize the
sky area lost while maximizing depth of exposure.

For stacked images of GRB fields we excluded the first snapshot of data from the stacked image as
the GRB tends to be bright at this time which would reduce the sensitivity to fainter sources in the image.

From this point onwards the process followed was the same for stacked images and single
observations\footnote{With the exception that, for stacked images the data preparation phase
is carried out for multiple event lists, once per observation in the image.}, and
the phrase `dataset' refers to either of these. 

\subsection{Data preparation}
\label{sec:dataprep}

Source detection was performed on a single image (in each band) which contained all of the usable (Section~\ref{sec:selection})
exposure time in that dataset. However the background maps had to be created on a per-snapshot basis and
then combined to give the full map (see Section~\ref{sec:bgmap} for details). The
datasets were therefore split into snapshots, and for each snapshot an exposure map was created (which included the effects 
of vignetting, assuming an event energy of 1.5 keV which is where the XRT effective area is at its highest) and an image was 
constructed of the grade 0-—12 events in each of the four energy bands (Table~\ref{tab:summary}).
The center of the image and the 
mean spacecraft roll angle for that snapshot were recorded to be used by the 
background-mapping software. The XRT has three different window sizes that 
have been used at different times: $480\times480$ pixels, $500\times500$ pixels, and 
$600\times600$ pixels; the size that was used was also recorded. Finally, the per-snapshot exposure maps 
were summed to give a single, total exposure map (as well as the 
per snapshot maps) as were the images in each energy band. These files were 
then passed to the source detection software.


\subsection{Source Detection}
\label{sec:sdet}
Source detection was performed independently for each energy band. We used
a form of sliding-cell detection combined with a fit to the point spread function (PSF) to identify,
localize and characterize sources. Our approach is based on that employed for the 2XMM 
catalog \citep{Watson09}, optimized for \swift-XRT data. The algorithm is composed of 
the following elements:

\begin{enumerate}
\item{Sliding-cell detection with a locally-estimated background}
\item{Creation of a background map}
\item{Sliding-cell detection using the background map}
\item{Source characterization using a PSF fit}
\item{Likelihood testing}
\end{enumerate}

The source detection process is non-linear and iterative. The specific details (e.g.\ thresholds) 
and ordering of the steps were optimized through a series of trials and simulations. An overview of
the algorithm is given in the rest of this section; the components of that algorithm
are detailed in the following sections.

A flow-chart depicting the source-detection algorithm is shown in Fig.~\ref{fig:procflowchart}.
The initial step was a sliding-cell detection with a locally estimated background. The source list thus produced
was required only to produce the initial background map and only needs to identify the brightest sources. We therefore used
a signal-to-noise (SNR) threshold of 10. Only the brightest source was considered, and this was only used to create
a background map, and then discarded. A second sliding-cell detection was then performed, this time using
the background map. The SNR threshold at this point was still 10, and only the single brightest source detected was kept.
This is necessary to avoid detecting artifacts around bright sources. If a source was detected at this point,
the PSF was fitted to the source and a likelihood test was performed. If the likelihood value (Section~\ref{sec:likely})
was below 3, the source was discarded as spurious; its position was noted so that, if the object were redetected in a later
step, it could be immediately discarded. The background map was then rebuilt, and the model PSF
of the detected source was added to the map, which reduces the probability of detecting the artifacts just alluded to
(see Section~\ref{sec:bgmap} for details). The siding-cell detection using the background map, and subsequent
steps, were then repeated until no new objects were detected. These steps correspond to the left-hand column
in  Fig.~\ref{fig:procflowchart}.

The detection threshold was then reduced to SNR=1.6 and the process continued largely as
above (build background map, detect, PSF fit; repeat) except that \emph{all} objects
detected were passed to the PSF fit, rather than just the brightest one. This was repeated until no new objects
were found. This stage is represented by the central column of Fig.~\ref{fig:procflowchart}.

The final stage of the process was to perform a new PSF fit and 
likelihood test for each object detected. This was needed because 
because the initial steps carried out above were done before all of the 
objects had been detected, so the background map will have evolved since 
this time. We therefore created the background map, adding in the model 
PSFs of all but the highest SNR object. We then performed the PSF fit 
and likelihood test on the highest SNR object, using this map. This 
process was then repeated with the second-highest SNR object left out of 
the map (the highest SNR source, relocalized in the previous iteration, is included)
and the PSF fit and likelihood tests performed for that source, 
and so on through each source. Finally, two definitive background maps 
were created and saved: one containing only the background, one also 
including the model PSF of every object detected. These steps are shown 
in the right-hand column of  Fig.~\ref{fig:procflowchart}.

We will now describe the five principle components of this process.

\subsubsection{Sliding-cell detection with a locally-estimated background}
\label{sec:cell_local}

Use of a locally-estimated background was made only once in our process.
During this phase the SNR threshold for a detection was 10. The algorithm
employed was that detailed in the \emph{Chandra Detect} Reference
Manual\footnote{http://asc.harvard.edu/ciao/download/doc/\newline
detect\_manual/cell\_theory.html}.
We used a $21\times21$ pixel (=$49.5\arcsec$) cell  and stepped it over 
the entire image in steps of 7 pixels. For each step, we measured the 
number of events, $C$ in the cell. The error was calculated according to 
the \cite{Gehrels86} formula:

\begin{equation}
\label{eq:errors}
\sigma_C = 1.0+\sqrt{C+0.75}
\end{equation}

\noindent which approximates the Poisson distribution better than $\sqrt{C}$ for low values of $C$.

We also measured the number of events, $T$ in a cell of size $51\times51$ pixels with the same central
position as the source. If the real number of background events in the inner cell is $B$, and the number
contributed by a source at the center of that cell is $S$, then:

\begin{equation}
C=\alpha S + B 
\end{equation}
\begin{equation}
T=\beta S + \left(\frac{b}{d}\right)^2 B 
\end{equation}

\noindent where $\alpha=0.814$ and $\beta=0.937$ are the fraction of source counts expected
in the inner and outer cell respectively, determined from the 
PSF of XRT \cite{Moretti07}; $d=21$ and $b=51$ pixels are the widths of the inner and 
outer cells.

Solving for $B$ and then $S$ gives the SNR in the inner cell:

\begin{equation}
{\rm SNR} = \frac{S}{\sigma_S} = \frac{C\left(b^2-d^2\right)d^{-2}-Q}{\sqrt{\sigma_C^2 \left(b^2-d^2\right)^2d^{-4}+\sigma_Q^2}}
\end{equation}

\noindent where 
\begin{equation}
Q=T-C
\end{equation}

This implicitly assumes that the exposure is constant across both the inner and outer cell,
which may not be true. Therefore to determine $Q$, we measured the number of counts that were in the outer cell
but not the inner cell, and then multiplied this by $E_d/E_q$; where $E_d$ is the mean exposure per
pixel in the inner cell, and $E_q$ is the same calculated for pixels in the outer cell but not the inner one.

The 21-pixel wide cell is not necessarily optimal. We 
therefore searched for any 21-pixel cell with an SNR$\ge$1, and then 
investigated such cells further, by creating a $17\times17$ pixel cell 
and stepping this around within the original 21-pixel cell (using an 
outer cell reduced in proportion). If one of these smaller cells had an 
SNR larger than was found in the 21-pixel cell, then its position and size were noted. 
The cell was then reduced to 15 pixels and stepped around inside the 
21-pixel parent cell as before. This continued for cells of size 11, 9 and 
7 pixels, with the cell always being moved in steps of $d/3$ pixels ($d$ is the width of the cell, the step size
is rounded when non-integer) until no cell with an SNR greater than that in the 21-pixel region was 
found. Then all of the cells which were noted during this process were 
compared. If any cells overlapped, only that with the highest SNR was 
kept. For each cell thus found, a barycenter was calculated (using only 
counts within that cell), and also the box size with the maximal SNR was 
determined. If this box had SNR$\ge10$ then it was saved as an `excess': a possible source.

Once the entire image had been searched in this way, any duplicate excesses were removed. If there were overlapping
excesses\footnote{Because the 21-pixel cells are moved in steps of 7 pixels and thus overlap.},
the mean box size and position is determined, weighted according to the number of events in each cell.
A barycenter was then calculated, and the overlap check repeated; this time where excesses overlap, only that
with the highest SNR was kept; the others were discarded. The final result of this process was
a unique list of excesses with SNR$\ge10$.

\subsubsection{Creating a background map}
\label{sec:bgmap}
The  above method assumes that there is at 
most a single source within the test cell; where multiple sources are 
close together this will therefore incorrectly estimate the background 
level. It also assumes that the cell is large enough to accurately 
sample the background and that this is invariant across the cell. These 
statements may be untrue.


We therefore produced background maps to accurately model the background 
across the detector and included in this map the sources which had already been 
detected. This process was repeated many times during the source 
detection process and it is pivotal to our method:  in 
Section~\ref{sec:verifybg} we demonstrate that it is reliable.

Even within a single observation, each snapshot covers a slightly different area of sky
because it follows a new slew to the target. If we created a background map based on the full
exposure, this would contain artifacts at the edges of the per-snapshot fields of view (particularly
if the background level varies between snapshots, for example due to thermal variations
in the passively-cooled XRT, \citealt{Kennea05}). We therefore constructed the background
map separately for each snapshot using the images and exposure maps created in Section~\ref{sec:dataprep},
and then summed these to create the per-image background map. 
The process, described below, makes a single-snapshot, single-band background map, and
was performed for each energy band and snapshot independently.

The first step was to create a detector mask. Initially all pixels in the 
mask were set to 1, then all pixels in the region of the excesses 
already identified were set to 0. The definition of `in the region of' 
depended on the details of the excess. For all but the first and last 
background maps created for a dataset, the list of excesses comprised a mixture of 
those returned by the most recent cell detect run and those which had been PSF-fitted.
For the former, the position and 
count rate were not well known, so the masking was approximate: 
the count rate of the excess was estimated based on the size of the cell in which 
the excess was detected and the standard XRT PSF, and pixels were masked out to the radius 
at which the count-rate dropped below $10^{-5}$ ct sec$^{-1}$ pixel$^{-1}$ (or a maximum radius of 150 pixels). For PSF-fitted 
excesses the best-fitting PSF profile and count rate were known, so the 
mask radius was that where the count rate fell to 10$^{-6}$ ct sec$^{-1}$ pixel$^{-1}$ (a 
typical background level for an XRT exposure), again with a maximum of 
150 pixels. If this process resulted in more than 80\% of the image being 
masked out, the mask radius was reduced by 5\% and the process 
reperformed; this was repeated until less than 80\%\ of the image was masked
(we set a maximum of 100 iterations, but this was never reached).

The mask was multiplied by the original image to create a masked 
image (referred to as a `Swiss-Cheese image' by \emph{Rosat} and 
\emph{XMM}; \citealt{Voges99,Watson09}), i.e.\ one where ideally all 
events from the detected sources have been removed. This image was 
divided by the exposure map\footnote{Pixels with zero exposure are set 
to zero.} and rebinned into a $3\times3$ grid, with the uncertainty in 
each bin also calculated according to equation (1). If a box contained no 
unmasked pixels, the value of that box was set by interpolation from the 
neighboring boxes. The central pixel of each box was set to the value 
determined for that  box, and the rest of the image was populated using 
bilinear interpolation from these nine values. The resultant image was 
then multiplied by the exposure map to give the background map. This 
process differs from the XMM approach of using spline interpolation over 
a finer grid than employed here, however that process tended to overfit 
the \swift\ background. The above approach of linear interpolation and a 
 $3\times3$ grid was arrived at through an extensive period of testing, 
and represents an excellent level of accuracy 
(Section~\ref{sec:verifybg}) for a modest number of parameters. The 
uncertainties were propagated through this process to give a background 
error map. 

Once the background had been modeled in this way, any excesses which had
been PSF-fitted in previous iterations were added to the background (and background error) map;
to reduce the number of spurious detections near to bright sources, and 
increase sensitivity to sources which are close together. This was done using the PSF 
profile from the PSF fit (Section~\ref{sec:psffit}) which has been 
modified to include the spokes caused by the shadow of the mirror support 
structure, and out-of-time events (see the Appendix).

\begin{figure*}
\begin{center}
\includegraphics[width=5cm]{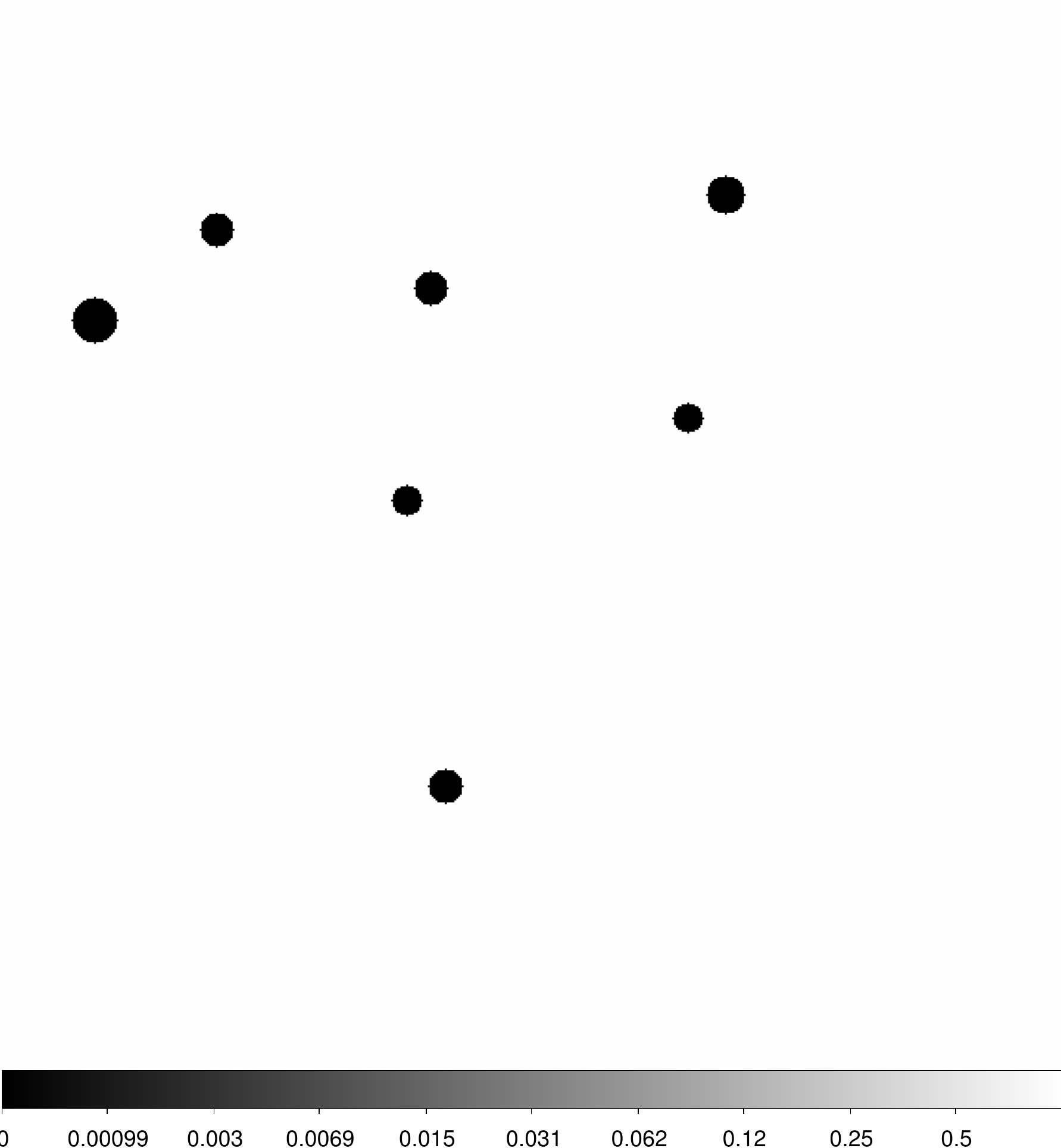}
\includegraphics[width=5cm]{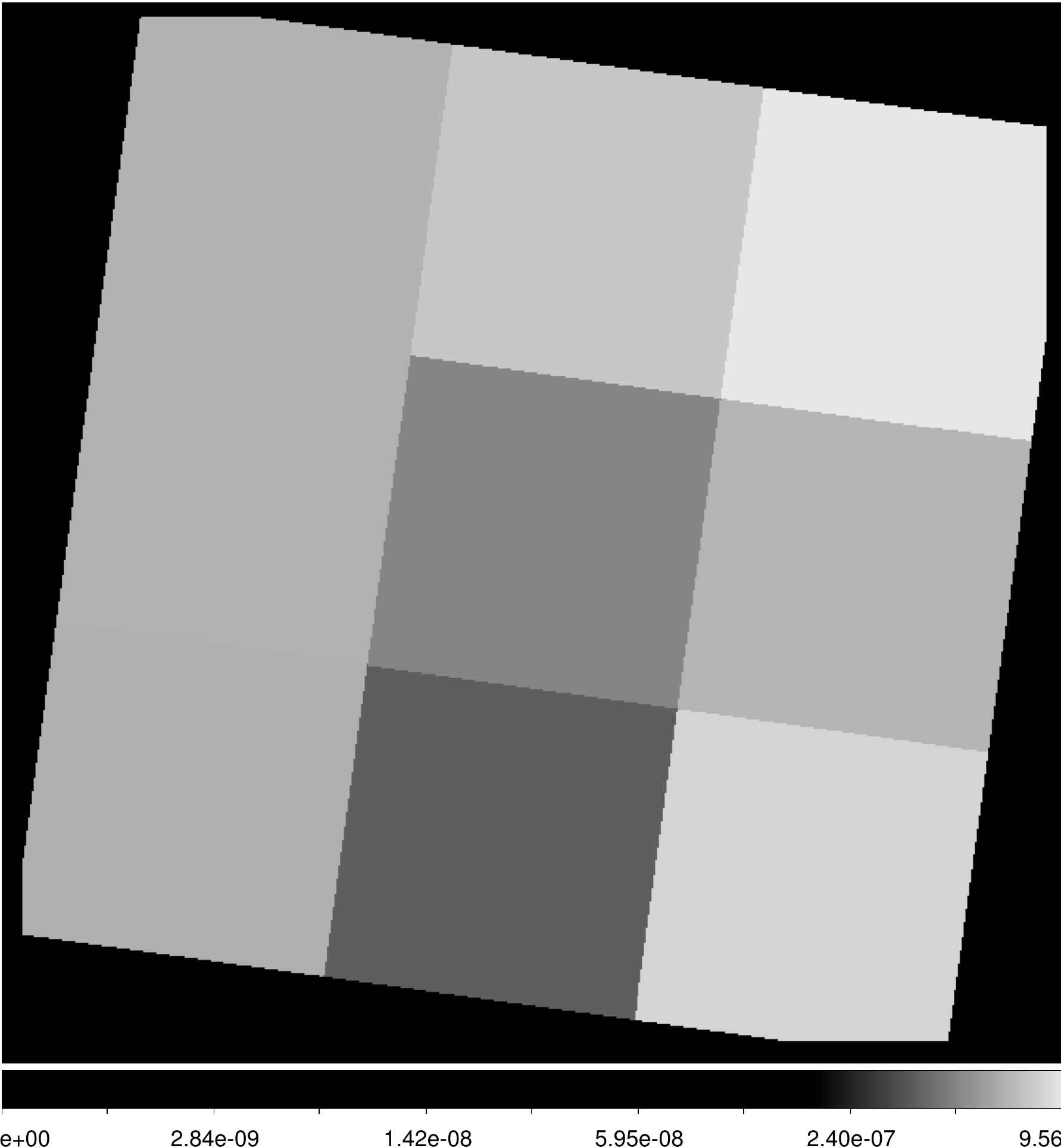}
\includegraphics[width=5cm]{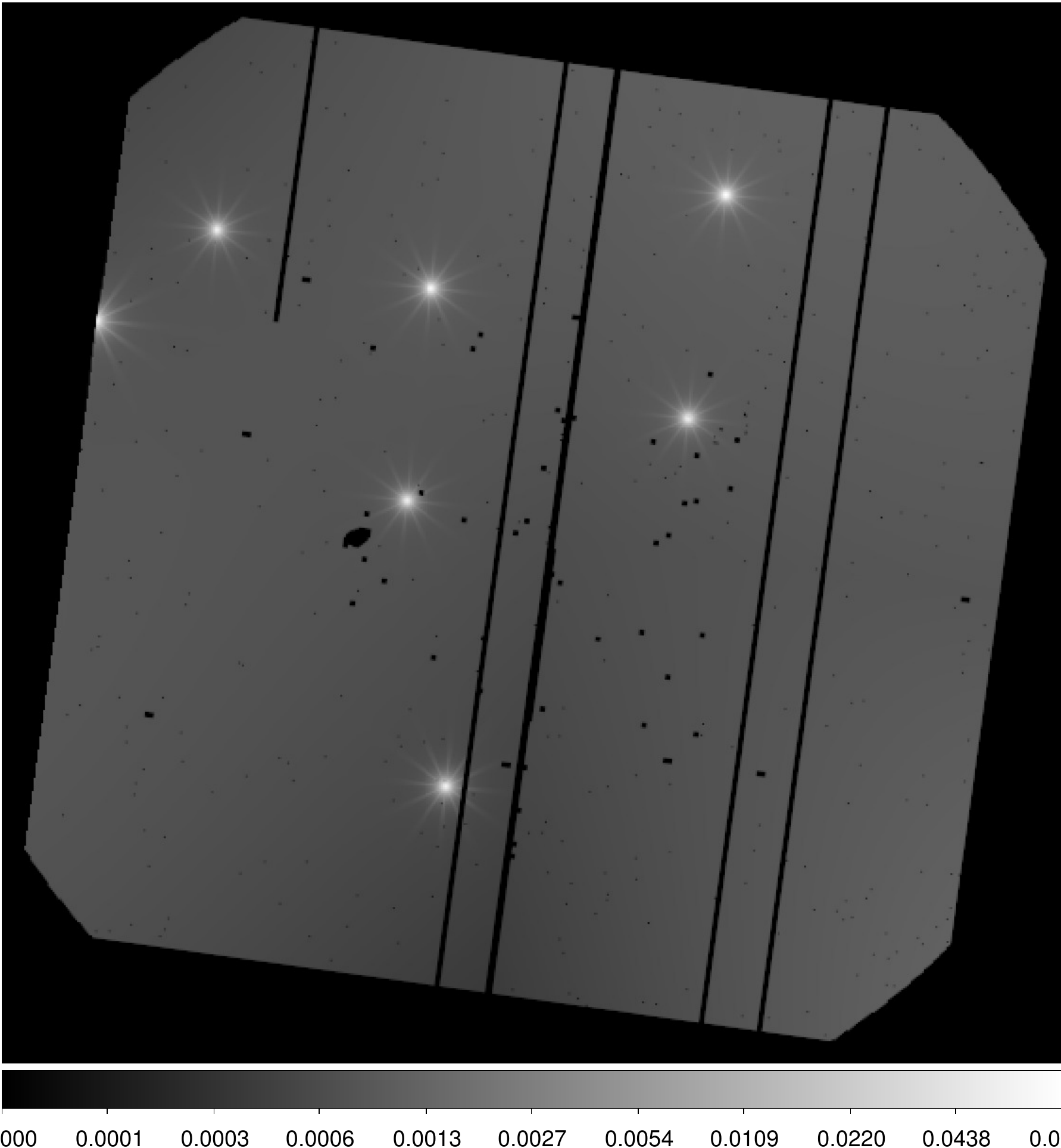}
\caption{Example stages of background map creation on a single snapshot. \emph{Left:} The detector mask; white pixels
are `on' while black ones are masked out. \emph{Center:} The rebinned background. \emph{Right:} The final background
map, including the model PSFs of the sources detected so far.}
\label{fig:bgmap}
\end{center}
\end{figure*}

The creation of a background map is illustrated in Fig.~\ref{fig:bgmap}.

\subsubsection{Sliding-cell detection using the background map}
\label{sec:cell_map}

This process was almost identical to that described in Section~\ref{sec:cell_local},
except that the outer cell was not used. Instead the background level, $B$ was simply
the sum of the background map within the cell, and hence

\begin{equation}
\label{eq:detmap}
{\rm SNR}= \frac{S}{\sigma_S} = \left(\frac{C-B}{\sqrt{\sigma_C^2 + \sigma_B^2}}\right)
\end{equation}

\noindent where $\sigma_C$ was defined as in Equation (1) and $\sigma_B$ was taken from
the background error map.

\subsubsection{Source characterization using a PSF fit}
\label{sec:psffit}

The positions of the excesses were determined using a PSF fit based
on that described in \cite{Goad07} and \cite{Evans09}. A circular region was selected, centered
on the position determined by the sliding-cell detection, with a radius based on the SNR
of the excess as given in Table~\ref{tab:psfrad}. The best position of the source was 
then determined by minimizing the C-stat \citep{Cash79} as modified for use in {\sc xspec}

\begin{equation}
\cstat= 2 \sum_i \left(M_i - D_i + D_i\left[\ln D_i - \ln M_i\right]\right)
\end{equation}

\noindent where the sum is over all pixels in the circular region, $D_i$ is the number of events
measured in a pixel $i$, and $M_i$ is the expected number of events in that pixel:

\begin{equation}
M_i= E_i \left(N P_i + B_i\right)
\end{equation}

\noindent where $E_i$ is the exposure, $N$ is the normalization, $P_i$ is the model PSF and $B$
the value of the background map, in pixel $i$.
We fitted for source position and normalization, using both the 
nominal PSF in the CALDB \citep{Moretti07} and the PSFs determined for piled-up sources \citep{Evans09}; these were
first modified to include the shadows of the telescope's mirror support
structure (see Appendix A for details).
Based on simulations, we required that \cstat\ decrease by at least 10 before accepting a more-piled-up
PSF as a better fit. Although the PSF is a function of both energy and off-axis angle, the
dependence on these factors is very small and we used the on-axis 1.5 keV profile for 
all of our fits. The 68\%\ confidence intervals on the RA and declination were determined independently,
by finding for each parameter the range of values within $\Delta \cstat=1$ of the best fitting \cstat\ value. This was later converted
to a 90\%\ confidence radial error via Rayleigh statistics, using $\sigma_{\rm Rayleigh}=0.5*(\sigma_x+\sigma_y)$.
For a small number of objects, the fit was unable to determine the uncertainty due to 
minimization errors. In these cases we set the 90\% confidence radial error 
to be $14.6\arcsec/\sqrt{N}$ (where $N$ is the number of events in the fitting region),
this relationship having been calibrated from simulations.

\begin{deluxetable}{cc}
\tablecaption{The radius of the region used to perform PSF fitting.}
\tablehead{
\colhead{SNR}      &       \colhead{Radius$^1$}  
}
\startdata
SNR $\le7$                   &    12 pixels \\
$7  < {\rm SNR} \le 11$  & 15 pixels \\
$11 < {\rm SNR} \le 40$  & 20 pixels \\
SNR $>40$                & 30 pixels \\
\enddata
\tablecomments{$^1$1 pixel=$2.357\arcsec$}
\label{tab:psfrad}
\end{deluxetable}

We then reconstructed the count-rate of the source, needed for the background map.
For most sources this was done using a circular region with radius
as for the PSF fit, but centered on the position returned by that fit.
However if the best-fitting PSF was one of the piled-up profiles,
or if the estimated count-rate in the original circle was $>$0.6 ct s$^{-1}$ (the level at which pile-up
tends to become significant) an annular
region was instead used, with the inner radius given in Table~\ref{tab:puprad}; these reflect the radii
at which the piled-up PSFs become asymptotic to the non-piled-up PSF. The outer radius was still that
used for the PSF fit if this was larger than the inner radius, otherwise it was 5 pixels more than that value.

The measured and background counts, $C$ and $B$, were taken from the image
and background map respectively in the region just defined. If $(C-B)>30$ then the estimated number
of source events, $S=C-B \pm \sqrt{C+B}$
as in Equation~\ref{eq:detmap} (except that we define $\sigma_C=\sqrt{C}$ as we are no longer in the low-count regime).
For lower numbers of measured counts the value $S$ was determined using
the Bayesian method of \cite{Kraft91}\footnote{At $C-B=30$ the Bayesian calculation converges with the standard
approach. However, the Bayesian approach assumes that there is no uncertainty in the 
background measurement which in principle leads to an underestimate of the error. For typical detections in our catalog this is
at the 0.1\%\ level, so can be ignored.}. To correct for the effects of pileup, vignetting
and exposure variations (e.g. due to dead columns on the CCD) we calculated the correction factor:

\begin{equation}
\kappa=P_{\rm inf}/P_{\rm meas}
\end{equation}

\noindent where

\begin{equation}
P_{\rm inf} = \sum_i^{r=150} \left( E_{im} P_i \right)
\end{equation}
\begin{equation}
P_{\rm meas} = \sum_i \left( E_{i} P_i \right)
\end{equation}

\noindent $P_{\rm inf}$ is the PSF summed from a radius of 0 out to 150 pixels (effectively infinity),
while $P_{\rm meas}$ was summed only over the region from which counts were measured. $E_{im}$ is the on-axis exposure of the image.
The estimated source count-rate is thus:
\begin{equation}
R = \frac{\kappa S}{E_{im}}
\end{equation}

\begin{deluxetable}{cc}
\tablecaption{The inner radius of the annular region used to measure the count-rate for piled up sources.}
\tablehead{
\colhead{Fitted PSF profile}      &       \colhead{Radius}  
}
\startdata
CALDB      & 3 pixels   \\
rate=0.9 ct s$^{-1}$  & 4 pixels \\
rate=1.4 ct s$^{-1}$  & 6 pixels \\
rate=2.6 ct s$^{-1}$  & 7 pixels \\
rate=4.0 ct s$^{-1}$  & 8 pixels \\
rate=5.2 ct s$^{-1}$  & 13 pixels \\
rate=8.6 ct s$^{-1}$  & 20 pixels \\
rate=15  ct s$^{-1}$  & 25 pixels \\
\enddata
\tablecomments{The `CALDB' profile is that determined by \cite{Moretti07} and given in the CALDB. The
remainder were determined by \cite{Evans09}. The `rate' is related to the object used to calibrate the PSF
and not to the source being characterized in this catalog. The PSF profile used to determine the count-rate correction factor
is the one determined in the PSF fitting stage.}
\label{tab:puprad}
\end{deluxetable}

We next checked for potential duplicates or detections of the same astrophysical object. These can occur in the PSF wings and diffraction 
spikes of bright sources, even though these were added to the background map
at each iteration. We therefore checked the distance of each newly-fitted excess from those found
in previous iterations. If it lay within the distance tabulated in Table~\ref{tab:mergerad} it was assumed to be an alias
of that object, and was discarded. 
This means that our detection method is blind to new sources in the close vicinity of brighter objects, however the tendency
to detect false positives in this region had effectively blinded the system anyway. Due to the nature of \swift's observing strategy,
this limit is often only temporary. For example, a newly detected GRB is usually bright, so the radius over which we cannot
detect new sources is large, however the GRB is observed again as it fades; in those later observations sources close to the GRB
can be reliably detected.

\begin{deluxetable}{cc}
\tablecaption{The distance from a source within which detections are assumed to be artifacts.}
\tablehead{
\colhead{Source rate }      &       \colhead{Radius}  \\
\colhead{(count/sec)}         &  \colhead{(pixels)}
} 
\startdata
$R\le0.4$  & 10   \\
$0.5<R\le1$ & 35 \\
$1<R\le2$ & 40 \\
$2<R\le8$ & 47 \\
$R>8$ & 70 \\
\enddata
\label{tab:mergerad}
\end{deluxetable}

\subsubsection{Detection likelihood}
\label{sec:likely}

After PSF-fitting an excess we calculated \cstat\ a second time with 
the normalization set to 0, i.e.\ with no source present. Since 
$\Delta \cstat$ is distributed as $\Delta\chi^2$ (\citealt{Cash79}; here with two degrees of 
freedom, $\nu=2$) we determined the probability that the change in 
fit statistic with and without a source present is coincidence: $P=\Gamma 
(\nu/2, \Delta \cstat/2)$ (where $\Gamma$ is the incomplete Gamma function), and the log-likelihood, $L=-\ln(P)$. As 
\cite{Watson09} pointed out, we cannot take this statistic at face value;
indeed the false positive levels they report are 10--100 times higher than
expected from the equations above for the likelihood values they quote.This is
because the  measurement with no source present is a boundary condition of the model: 
as the source normalization cannot be negative, the test with normalization set to 0 is at the limit of 
the allowable model space. In such cases the likelihood ratio does not follow a \chisq\ distribution (see \citealt{Protassov02} for
a detailed discussion).  Like \cite{Watson09} we instead calibrated 
the relationship between $L$ and $P_{\rm false}$ using simulations, as described 
in the Section~\ref{sec:verify}. Based on this calibration, we rejected any excess with $L<3$.


\subsection{Quality flags and further checks}
\label{sec:check}
\label{sec:ol}

Several further tests were performed to eliminate spurious or extended sources and 
to indicate how reliable a given detection is. Spurious detections can arise due to 
hot columns and hot rows on the detector. For each excess, we selected from the relevant event 
list all the events lying within the PSF fitted region. 
Only excesses containing events from at least three distinct detector pixels, rows and columns were accepted;
in addition, any excess where
$>$50\%\ of the  events lie in a single pixel, or $>$75\%\ lie within a single row or column was discarded.
After this the location of each surviving excess was compared
to a list of known extended objects (taken from \citealt{Tundo12}): if the excess lay within
the extent of the extended object it was discarded. 

The remaining excesses are considered to be detections of genuine astrophysical sources, but some level of 
contamination will remain: we therefore assigned each source a quality flag to indicate
the probability that it is a false positive. This flag is a function of the 
exposure time and the likelihood value for the source, and can be either \good, \reasonable\/\ or \poor\ 
(with corresponding integer values of 0, 1 and 2). If only \good\ sources are considered, the false positive
rate is 0.3\%; if \good\ and \reasonable\ sources
are included, this rises to 1\%, and if \poor\ sources are also considered, the false positive rate is 10\%.
Of course the fraction of true sources that are detected (i.e.\ the completeness) also rises as \reasonable\ and 
\poor\ detections are included. This allows users to easily choose between sample size and sample purity. Full details of
the definitions of the quality flags and how the false positive rate and completeness fraction were calibrated are given in Section~\ref{sec:falsep}.

There is an additional category of sources, \emph{Bad}, which is not 
included in our catalog. Such sources were accepted by the source 
detection code, but as they have a very high false positive rate (\til 
80\%) they were rejected before the detections are merged 
(Section~\ref{sec:obssource}). The background map was reconstructed at 
this point without the \emph{Bad} detections considered. This new 
background map was used for construction of the source count rates and 
light curves. We stored a list of these \emph{Bad} detections for use 
with the upper limit server (Section~\ref{sec:upperlimit}).

We also performed an automated check for the phenomenon called \emph{optical loading}.
Bright optical sources can liberate sufficient charge in the XRT CCD because of
the large number of optical photons accumulated in a 2.5-s PC mode 
exposure frame that the characteristics of  X-ray events at the 
location of the optical source are distorted. When this first becomes a 
problem, it causes the energy of the X-ray events to be overestimated\footnote{A correction for this is made by the {\sc xrtpccorr} tool called by {\sc xrtpipeline} as part of the standard processing.}. 
At higher optical fluxes, it can cause real X-ray events to be discarded 
or spurious events to be detected. The flux at which this occurs is a 
function of stellar color and is discussed in detail at 
http://www.swift.ac.uk/analysis/xrt/optical\_loading.php; stars brighter than 
$V$\til9 can be a problem, the limit being more severe for those later than $M0$.
We set a threshold at which optical loading is to be flagged as that at which a 
star contributes spurious events at a level of $\ge 10^{-3}$ ct 
s$^{-1}$. We searched for cataloged stars above this threshold within 
30\arcsec\ of each X-ray source in our catalog, using their cataloged $B-V$ color
to estimate spectral type and hence determine the $V$ magnitude limit.
If such a star was 
found, a field {\sc ol\_warn} is set in the catalog, indicating how many 
magnitudes brighter than the threshold the star is. We used the Tycho-2 
\citep{Hog00}, Bright Star Catalog \citep{Warren87} and General 
Catalog Of Variable Stars \citep{Samus10} as our source of optical 
objects. These sometimes contain the peak magnitude of a variable object,
which may not be appropriate to the \swift\ observations  (e.g.\ GK~Perseii
has a catalog magnitude of \til0, based on its nova eruption of 1908; but was at least 10 magnitudes fainter during all \swift\ observations),
so a large {\sc ol\_warn} value should be taken as a warning that an object 
\emph{may} be affected by optical loading, rather than that it is affected.

\subsection{Merging detections across bands}
\label{sec:obssource}

Since the detection system was performed independently on the four energy bands within a dataset,
the list of sources detected in each band had to be merged to create a unique list of sources for that dataset.
This was done by considering the detected sources in descending order of SNR and then using the radii given
in Table~\ref{tab:mergerad} to determine which detections correspond to the same object. Where a source was detected
in multiple energy bands, the definitive position of that source (in this dataset) was
taken from the detection with the smallest position error 
(provided this is not one where the error could not be determined from the fit).

For a source which was undetected in one or more energy bands, the images of those bands
were examined to determine the number of events at the source location. The expected background
level was determined from the corresponding background map. The count-rate and error
for this energy band was then estimated using the Bayesian approach  of \cite{Kraft91} and the PSF correction $\kappa$
was applied as for detected sources.  Although the source was undetected
in this case, we did not produce an upper limit, even though the count-rate may well be consistent with zero. Instead
we determined the value and the 68\% (i.e.\ 1-$\sigma$) confidence limits, as we do for detections. Note that we
give the positive and negative uncertainties separately as, when using the Bayesian approach, they may not be the same.

We also determined two hardness ratios, defined as 

\begin{equation}
\label{eq:HR1}
{\rm HR1} = (M-S)/(M+S)
\end{equation}

\begin{equation}
\label{eq:HR2}
{\rm HR2} = (H-M)/(H+M)
\end{equation}

Where $S, M, H$ refer to the soft, medium and hard bands respectively. If both bands in the hardness ratio
contained $>100$ counts, and had a SNR$>2$ then the ratios were calculated using the above equations, with the 
errors on $H$, $S$ and $M$ taken as $\sqrt\{H,M,S\}$ respectively and propagated through
equations~\ref{eq:HR1} and \ref{eq:HR2}. For fainter sources we used the Bayesian method of \cite{Park06}, where we used the effective area option in their
code to include the count-rate correction factors in the calculation. While the Bayesian method gives asymmetric errors (which are typically 
a few percent larger than the standard method returns), the standard method returns symmetric errors. This means one can find, for example,
HR1=$0.95\pm0.1$, even though the HR must be between $-1$ and 1 (inclusive). In such cases of course, the true HR limit is +1 (or -1 in a negative
counterexample).

\subsection{Manual screening}
\label{sec:visscreen}

\begin{figure}
\begin{center}
\includegraphics[width=8.1cm]{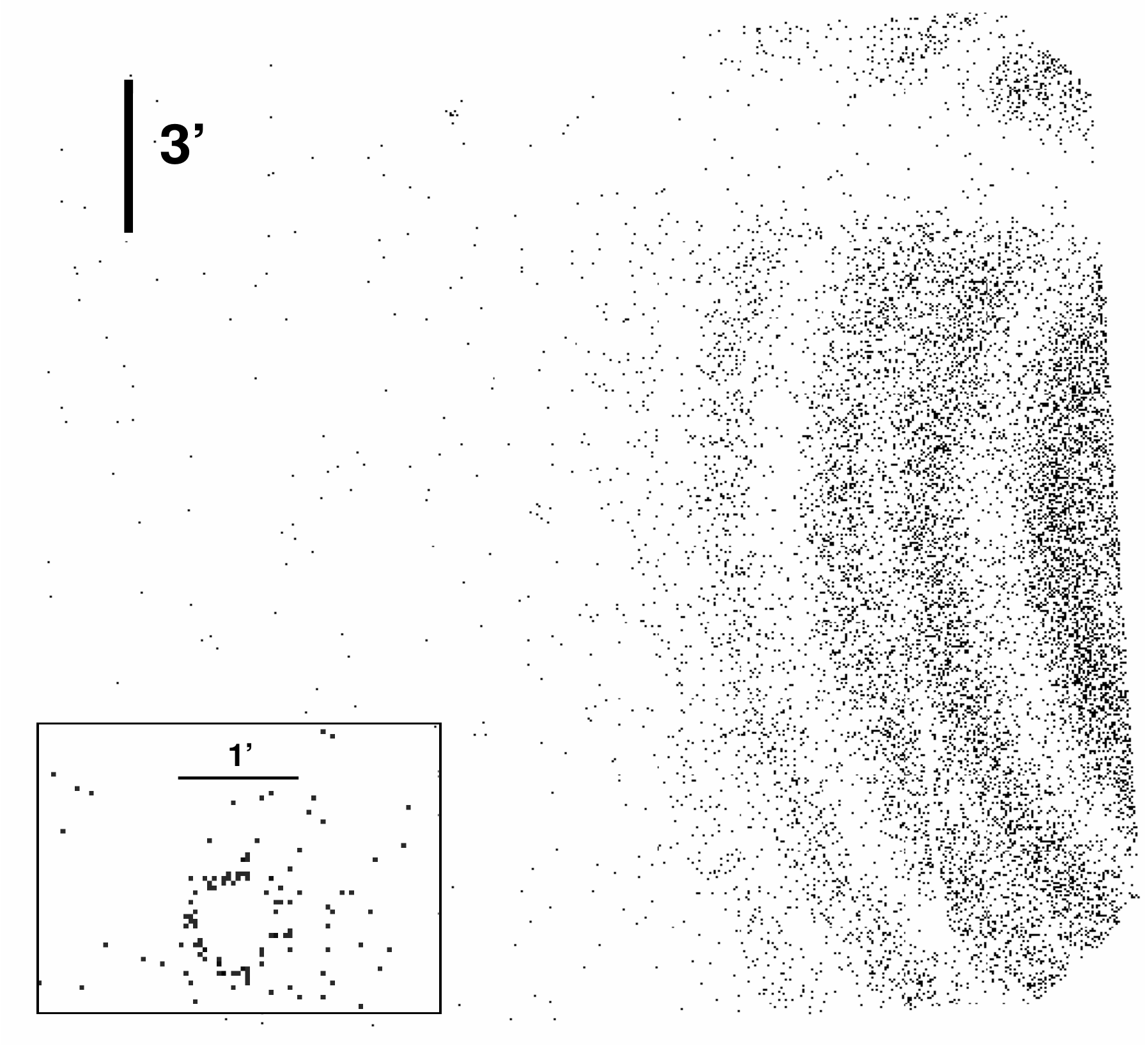}
\caption{Examples of artifacts that were identified by manual screening. 
The main plot shows \emph{stray light}: caused by
single reflections from a bright source lying outside the XRT field of view -- in this case the Crab nebula,
lying 45\arcmin off-axis. The gap in the rings is the shadow of the mirror support structure. 
\emph{Inset:} a `ring of fire': the apparent X-ray events were 
caused by accumulation of optical photons from a bright star (a $V=3$ Be star in this example).
Towards the center of the star's location no
events are detected because the optical flux is so high that in a single 2.5-s CCD exposure frame all pixels 
register events, and thus the event `grade' (which describes how many pixels a given event affected) is above
the maximum value permitted for valid events.}
\label{fig:sl}
\end{center}
\end{figure}

While the quality flagging system based on the source likelihood values 
is reliable for celestial point sources, it can be deceived in the presence of 
structured diffuse emission (e.g.\ from a supernova remnant) or 
instrumental artifacts. The most common of these artifacts is
\emph{stray light} \citep{Moretti09}: X-ray photons from a source 35\arcmin--75\arcmin\ off-axis (i.e.\ 
outside the field of view) that are directed onto the XRT detector 
via a single reflection (as opposed to the double reflection which focuses X-rays).
This occurs at a very low level: the effective 
area of the XRT for a source 50\arcmin\ off-axis is \til33,000 times 
lower than that on axis, and the singly-reflected photons 
are distributed over a much wider area of the CCD than for a focused source. Nonetheless, sufficiently 
bright sources outside the field of view can cause concentric arcs of events to be detected 
in the CCD (Fig.~\ref{fig:sl}) which can give rise to spurious source
detections. 

The typical background level of the observations in our 
catalog is \til$10^{-6}$ ct sec$^{-1}$ pixel$^{-1}$; for a source outside the
field of view to contribute stray light at this level it would require an 
on-axis XRT count-rate of \til 3 ct sec$^{-1}$. We conservatively chose a 
limit of 1 ct sec$^{-1}$, which corresponds to a 0.3--10 keV flux of 3.5\tim{-11} erg \cms 
s$^{-1}$ assuming a typical AGN spectrum: a power-law spectrum with $N_H=3\tim{20}$ \cms and 
$\Gamma$=1.7. We identified all sources in the 
\emph{Rosat} PSPC and 2XMMi-DR3 catalogs with fluxes above this limit, and selected for manual screening all fields in our catalog 
that lay within 28\arcmin--82\arcmin\ of those sources. This did not identify all
fields affected by stray light, as 2XMMiDR3 covers only a small 
fraction of the sky, \emph{Rosat} is not sensitive to strongly 
absorbed or hard sources and some objects are variable.

There are other
artifacts that can contaminate the images. These are residual bright Earth contamination, the `ring of fire' effect caused by 
serious optical loading (Fig.~\ref{fig:sl}, bottom) and the presence of extended 
sources or diffuse emission. All of these effects (and stray light) give 
rise to spatially proximate spurious detections. For this reason we also selected for manual screening any image where 
the median distance between detections was $<80\arcsec$. In total 15,152 
datasets (out of 56,275 in the catalog) were selected for human 
inspection.

We inspected these images in decreasing order of exposure time. If an 
image was deemed to be affected by the artifacts described above, then 
the results of this screening was applied to all pointings covering that 
location on the sky, avoiding the need to check each image individually.
When artifacts were manually identified, regions 
were defined which encompassed them, and any sources which lay 
within those regions had their detection flags changed. The `Field 
flag' for the image was also  set from its default value of \good (=0) 
to \emph{Flagged} (=1 or 2). For images containing artifacts (stray 
light, bright Earth or rings of fire) the detection flag of affected 
sources was increased (from 0,1 or 2) by 8 and the field flag set to 1. For images 
containing diffuse emission the detection flag of affected sources was 
increased by 16 and the field flag set to 2.

We distinguish between artifacts and diffuse emission because, while both of these phenomena
affect the background map (by causing inhomogeneities over
which the background map attempts to smooth and interpolate, and potentially by causing the detection
of spurious sources which in turn are added into the background map), artifacts
have well defined edges, but it is often not clear where a diffuse source stops contributing to the background.
For this reason (given that a dataset can only have a single field flag value) where both
artifacts and diffuse emission were identified in an image, the flag was set for the latter.

The result of the screening is that any source with a detection flag 
with a value $\ge8$ (i.e.\ lying inside a \emph{region} which has been manually 
marked as contaminated) has a high probability of being spurious, whereas sources 
with a flag value below this but lying in a \emph{Flagged} field (i.e.\ 
in the field, but outside the region manually marked as bad) have false 
positive rates as described in Section~\ref{sec:check}, but may have 
incorrect background values  and thus measured 
source fluxes.


\subsection{Astrometric corrections}
\label{sec:xastrom}

We attempted to derive a more accurate astrometric solution for our datasets than 
that available from the star trackers mounted on the XRT. The latter gives positions
accurate to 3.5\arcsec\ 90\% of the time \citep{Moretti07}. For each dataset, we matched 
the \good\ and \reasonable\ sources with the  2MASS catalog \citep{Skrutskie06} using an 
approach similar to that employed by \cite{Butler07d}. For every dataset in which more than two X-ray
sources were detected, we retrieved a list of 2MASS objects that lay within the XRT field of view
and attempted to find an aspect solution for the field which maximized the likelihood:

\begin{equation}
L=\sum_{{\rm ox}<20} e^{\left(-0.5\delta^2/\sigma^2\right)}
\end{equation}

\noindent where $ox$ is the angular separation between each \good\ and \reasonable\ XRT source and each 2MASS source,
so the sum is over all XRT/2MASS source pairs  within 20\arcsec\ of each 
other; $\delta$  is the angular distance between the 2MASS and XRT sources in question, and  $\sigma$ is the radial 
uncertainty in the two positions added in quadrature. The 1-$\sigma$ uncertainty in  the aspect
solution thus derived was taken as the RMS of the $\delta$ value for each 2MASS/XRT pair in the final fit.
If the mean shift in any of the X-ray positions as a result of this process was $>15\arcsec$ then the solution
was considered unreliable and rejected: this distance corresponds to a 7-$\sigma$ inaccuracy in the star tracker solution, which is a most unlikely
situation.

This process could not find an astrometric solution for every dataset, and in the majority of cases where a solution was found, the uncertainty
in the aspect solution was $>3.5\arcsec$; in these cases we used the star tracker attitude. A solution with an error $<3.5\arcsec$ was
found for only 4\% of the datasets in our catalog, but as these were the datasets with
objects in them, 26\%\ of the sources in our final catalog have positions improved using this technique.
Whichever method was used, the astrometric error was added in quadrature to the statistical position error
from the PSF fit (Section~\ref{sec:psffit}) to give the radial position error reported in the catalog.

To verify that this method gives reliable positions and uncertainties, we applied it to the fields containing the 999
objects in our catalog which are within 20\arcsec of quasars in the SDSS Quasar Catalog DR5 \citep{Schneider07} and thus likely to be
the X-ray counterpart to the quasar. We found that  90\% of the XRT positions thus produced agreed with the SDSS positions at the 90\%\ confidence level, 
as expected.

\subsection{Building the final unique source list}
\label{sec:sourcelist}
Once the above steps had been completed for every dataset contributing to the catalog, we 
merged the lists of sources from each dataset into one final source list.
This was done in the same way as described in Section~\ref{sec:obssource},
except that instead of using fixed merge radii based on the source brightness,
different detections were assumed to be the same source if their positions agreed at the 
99.99999426\% level\footnote{\label{note:rayleigh}i.e.\ the 5-$\sigma$ level of a Gaussian distribution. Since radial
errors follow a Rayleigh distribution, we use the probability level, not the number of $\sigma$. By `the positions
agreed' at this level we mean that the probability from Rayleigh statistics of their separation being that observed
or lower, given their position errors, is less than this threshold.}. For a typical 
source this was \til14\arcsec (\til 6 XRT pixels), which is 75\%\ of the PSF FWHM and the probability
of distinct sources lying this close to each other is very low, $\ll1\%$.

This approach takes into account the fact that different observations may have different astrometric accuracy, and allows
for faint sources that are near to a bright source, but not detected until after that object has faded, to
be distinguished from the bright source. When compiling this final source list, the detection flag in each band
was set to the best of the detection flags in that band from the individual detections of the source. A final, overall detection
flag was also produced which was the best of the per-band flags, and likewise for the field flag. The optical loading warning
was set to be the worst value from the set of individual detections of the source. The final source position was 
taken from the detection with the smallest position error, and the source was given a unique designation of the form:
1SXPS JHHMMSS.S+DDMMSS. This acronym has been registered with the IAU.

\begin{figure}
\begin{center}
\includegraphics[width=8.1cm]{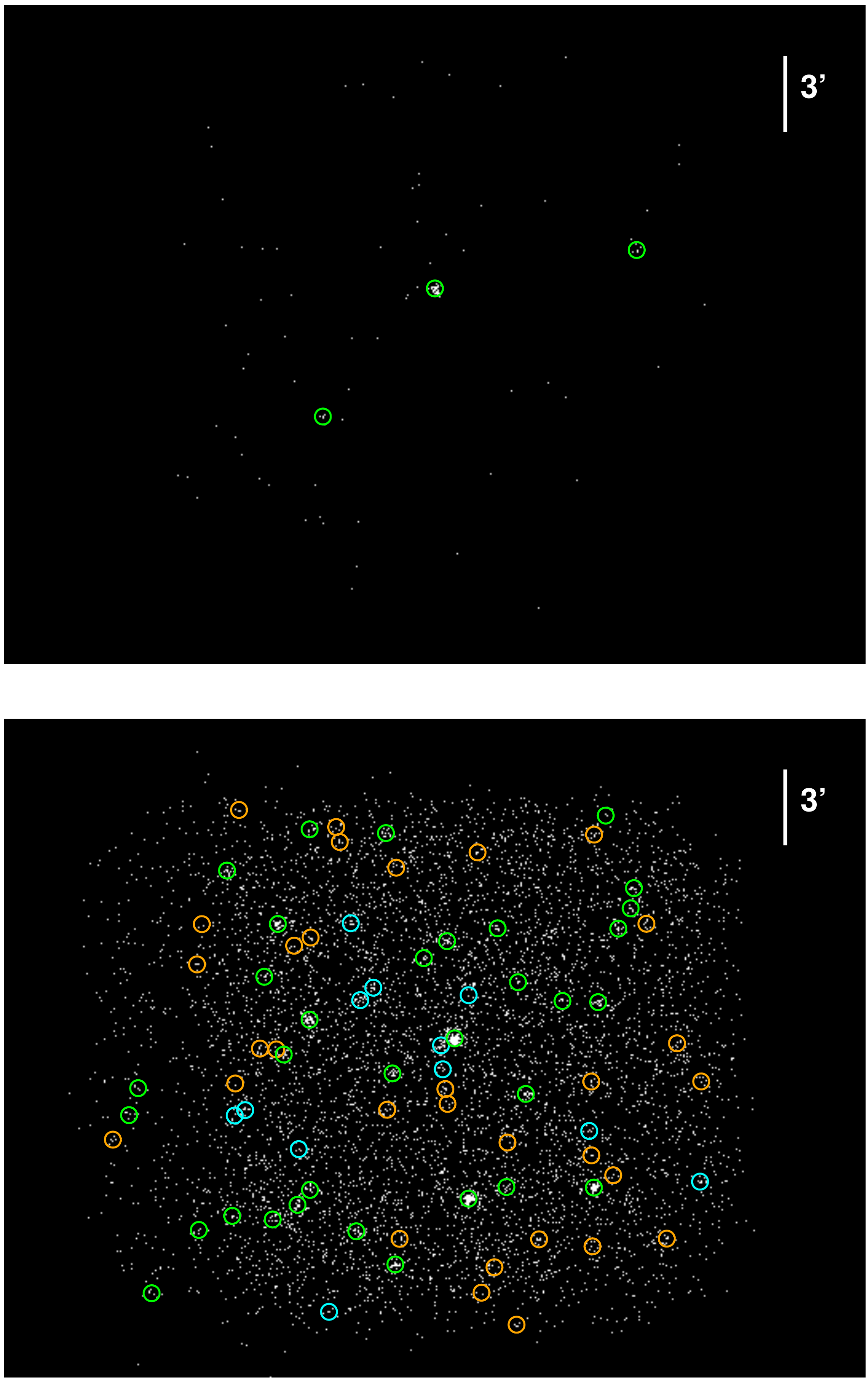}
\caption{Example datasets from the catalog. Both images are from the total band (0.3--10 keV) 
with pixel intensity following a log scale. \emph{Top:} a short single observation,
(ObsID 00032165001, exposure 424 s) with the three sources detected in that observation and band shown.
\emph{Bottom:} a deep stacked image (field 7086, exposure 1.1 Ms); the final unique source list for
this region is shown. The regions indicate objects detected, with 
the `quality' of the detection shown by the color: green=\good, cyan=\reasonable, orange=\poor. The regions
are a fixed size and do not reflect the size of the region used in source detection.
}
\label{fig:example_im}
\end{center}
\end{figure}

Fig.~\ref{fig:example_im} shows two examples of datasets after all of the steps in this section have been applied.

\section{Source-specific products} 
\label{sec:prods}

The details of the unique sources and the individual detections are available in the 
form of catalog tables, available to query online or and download (Section~\ref{sec:about}). 
In addition to these, we have produced light curves, hardness 
ratios and  variability and flux estimates for each source, and spectra 
for the brightest sources, as described below. These products are 
available to download via the 1SXPS website, where tools also exist to
calculate upper limits for specific locations on the sky.

\subsection{Temporal products}
\label{sec:temporal}

We produced light curves in each of the four energy bands, with one bin per observation and one
bin per snapshot (for observations where the source is undetected the latter light curve
only contains a single bin integrated over that observation). We also produced time series of the
hardness ratios with one bin per observation. The times of each bin in all of these products were corrected to 
the solar system barycenter (i.e.\ TDB).

To construct the time series, the count rate in each snapshot or observation was determined as 
described in Section~\ref{sec:psffit}, except that we used the best source position determined 
per observation (see Section~\ref{sec:obssource}), to account for the potential differences in astrometry between observations. The 
source-count accumulation region used was also that from Section~\ref{sec:obssource} if the source was detected; for bands, snapshots or
observations where the source was not detected a circular region of radius 12 pixels (28.3\arcsec) was used.

For the time series in each band we calculated used the Pearson's \chisq\ \citep{Pearson00}
to determine the probability that the source was variable.
The Pearson's \chisq\ is defined as:
\begin{equation}
\chi^2=\sum_i{\left\{\frac{\left(D_i-M_i\right)^2}{M_i}\right\}}
\end{equation}

\noindent where $D$ and $M$ are the data and model in bin $i$ respectively.  
These must be not in units of the count-rate (as contained in the light curve), but the measured number of counts ($C$) in each bin.
Since we test for the null hypothesis that
the source is constant, the model is that of constant source flux, but this is not the same
as constant source counts in each bin as the exposure time ($E$) and
count-rate correction factor ($\kappa$, see Section~\ref{sec:psffit}) can vary from bin to bin. Explicitly 
including these factors and the background level, if the source is constant the count rate is the same in each bin and is simply the
mean value; which can be determined from the measurements thus:

\begin{equation}
\begin{split}
R_i  & = {\rm const} \\
	 & =  \frac{(C_i-B_i)\kappa_i}{E_i}  \\   
     & = \frac{ \sum_j{\left\{\kappa_j\left(C_j-B_j\right)\right\}}   }{E_{\rm tot}}
\end{split}
\end{equation}

\noindent where the summation is over all bins, and gives the total number of PSF-corrected counts over the light curve.
We can then solve the above to determine the model of the number of counts per bin:

\begin{equation}
\begin{split}
M_i  & =  C_i \\
     & =  \left(\frac{E_i}{E_{\rm tot}}\right) \left(\frac{\sum_j{\left\{\kappa_j\left(C_j-B_j\right)\right\}}}{\kappa_i}\right) + B_i 
\end{split}
\end{equation}

This test, which reports the probability of the null hypothesis that the source is constant,
was applied to both the per-snapshot and per-observation 
light curves (but not the hardness ratio time series), probing variability on multiple timescales. To ensure that the per-snapshot result
is not affected by variation on the per-observation timescale, we calculated \chisq\ and hence $P$ for the per-snapshot
light curve of each observation independently, and then report the lowest value thus obtained.

We also tried using the Wald-Wolfowitz runs test \citep{Wald40} as an independent
measure of variability, however this lacked the power to identify variable sources
in our catalog, probably because many light curves have small numbers of bins. We therefore
elected not to include these results in the catalog.

\subsection{Flux conversions and spectra.}
\label{sec:spec}

\begin{figure}
\begin{center}
\includegraphics[width=8.1cm]{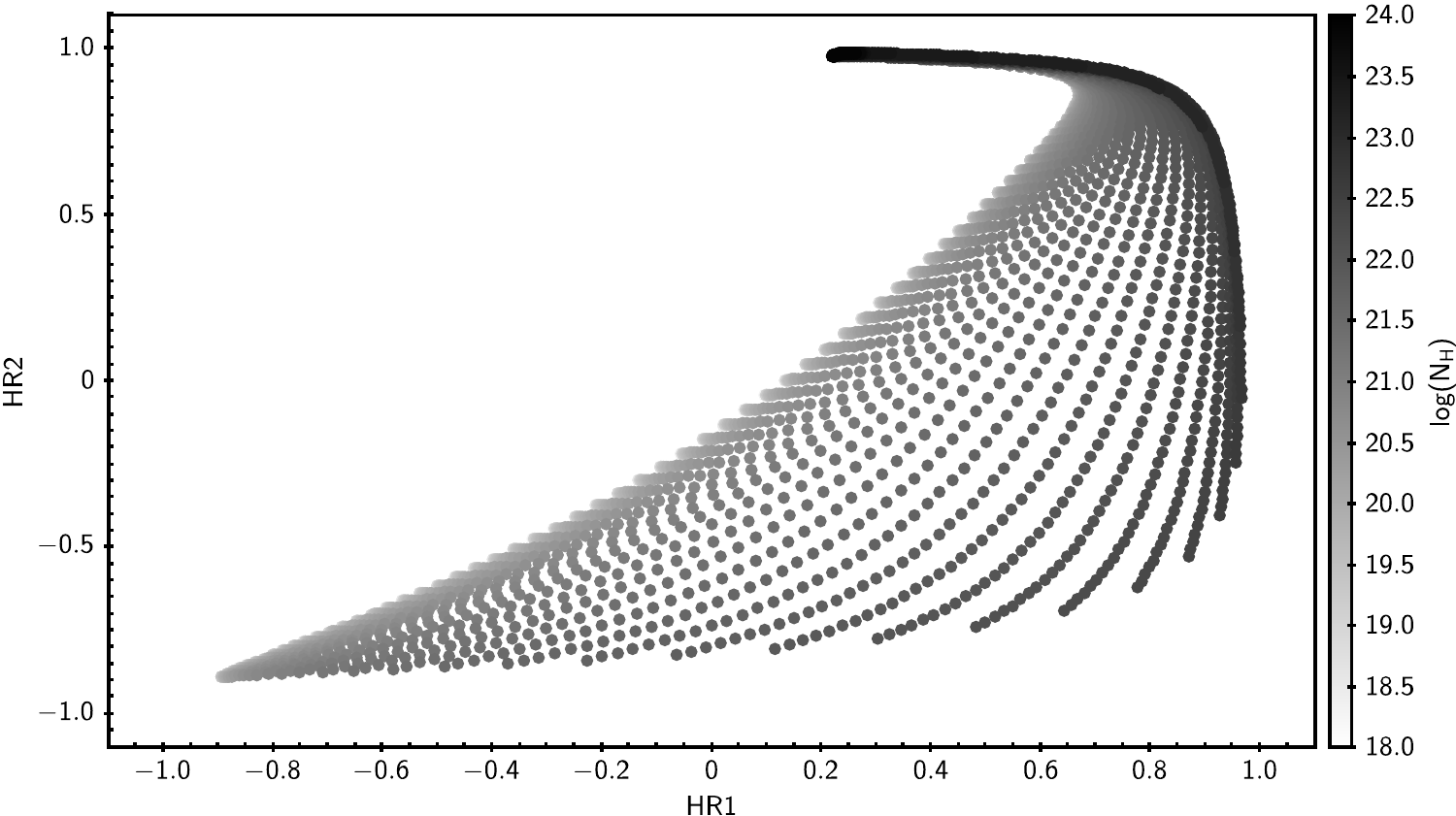}
\includegraphics[width=8.1cm]{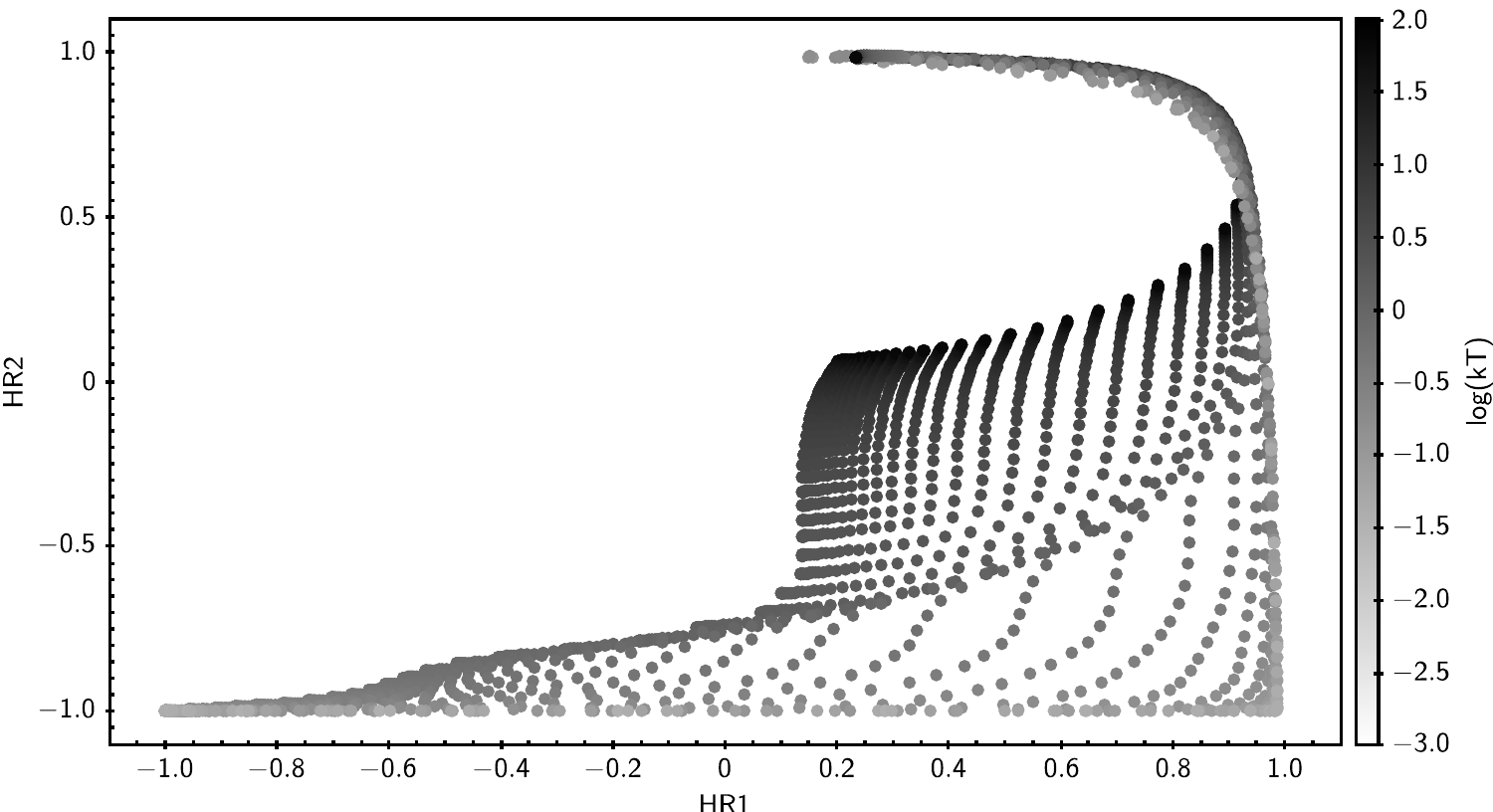}
\caption{The (HR1,HR2) plots used to deduce spectral information for the sources.
\emph{Top:} For a power-law spectrum, the $N_H$ values (grayscale) as a function of (HR1,HR2); each point
also has a $\Gamma$ value and ECF, not shown here.
\emph{Bottom:} For an APEC spectrum, the $kT$ values are shown as the greyscale. Each point also has $N_H$ and ECF values
not shown here.
}
\label{fig:HRinterp}
\end{center}
\end{figure}

For every source in the catalog we created energy conversion factors (ECF) to 
convert from count-rate to flux (observed and unabsorbed)\footnote{The uncertainty in the ECF
was not propagated into the error on the flux; this was simply the count-rate error multiplied by the ECF.}
We did this for two commonly observed spectral types: an absorbed power-law
and an absorbed APEC optically thin thermal plasma model
\citep{Smith01}; for the latter we assumed solar abundances.
 The absorption was modeled using the {\sc tbabs} model \citep{Wilms00}. 

For each source in the catalog we first determined ECFs using standardized spectra:
a power-law with a photon index of 1.7, and an APEC with a temperature of 1 keV; the absorption
was fixed at the Galactic value in the direction of the source, determined using the {\sc nhtot} tool of \cite{Willingale13}.

We also estimated the flux and spectral parameters from the hardness ratio information. Using {\sc xspec} we
simulated a series of spectra, with $17 \le \log N_H/(1\ \cms) \le 24$; for the power-law spectrum
we used photon indices in the range $-3\le\Gamma\le5$ and for the APEC spectrum we used
temperatures $-2\le \log (kT/{\rm1 keV}) \le 1.9$. We folded each simulated spectrum through the instrument response to derive 
its ECF and its two hardness ratios. We used the latter to construct a look-up table of the spectral parameters as a function of
(HR1, HR2); examples are given in Fig.~\ref{fig:HRinterp}. For each source in our catalog, if (HR1,HR2) lay 
in the region covered by the simulated spectra we interpolated on this grid to ascertain the spectral parameters
of the source. We also did this for the four points given by (HR1$\pm\sigma_{\rm HR1}$,HR2$\pm\sigma_{\rm HR2}$) to estimate the uncertainty 
on these properties. For any of those limits which lay outside the range covered by the simulated spectra, we took the values for the (HR1,HR2)
point nearest to the limit in question.  Note that the range of parameters for the simulated spectra goes beyond
what we may physically expect for XRT sources, in such extreme cases the purpose of this
approach is to give reasonable flux estimates within the 0.3--10 keV band, over which the model gives
an acceptable approximation to the data. However the actual the spectral parameters themselves should be viewed with caution
in those cases, and care should be used before extrapolating outside of the XRT bandpass.

For sources where (HR1, HR2) lay outside the range covered by the simulated spectra
we cannot calculate the spectral parameters in this way, instead we determined the probability of measuring (HR1,HR2)
if the true spectrum were that of the simulated spectrum with hardness
ratios closest to the measured values, given the uncertainties on those values.

For the sources with at least 50 net events in the total band,
we also built spectra using the software of \cite{Evans09}. We fitted these with an absorbed power-law and absorbed APEC,
with all parameters unconstrained (i.e.\ the fitted absorption was independent of the expected Galactic value).
The fit was performed on spectra binned to at least one photon per bin (i.e. {\sc group min 1} in {\sc grppha}), fitted using the {\sc xspec} 
$W$-statistic\footnote{i.e.\ by requesting the \cstat-statistic and then providing a background spectrum with Poisson statistics,
see https://heasarc.gsfc.nasa.gov/xanadu/xspec/manual/XSappendixStatistics.html};
after fitting we calculated \chisq\ using the Churazov weighting option \citep{Churazov96} to indicate the fit quality and allow
users to reject poor fits. Note that this is not a reliable goodness-of-fit indicator (see \citealt{Churazov96}, section 3.2) and cannot be used to calculate
the null hypothesis probability. 


In the final catalog table we report the spectral properties derived through all three of the above methods
(fixed spectra, interpolation of the HR values, and spectral fitting) where they are available. Since not all 
objects have all of the properties, this can make comparison of sources awkward, we have therefore 
included in the catalog a set of `best spectral properties'. These are taken from the spectral fit if it exists,
otherwise the HR interpolation, and if neither of those is available, the results from the fixed spectrum are used here.

\subsection{Cross correlation with external catalogs}
\label{sec:xcorr}

We cross-correlated the 1SXPS catalog with various external catalogs and 
databases, defining a source match to be where the 1SXPS and external 
catalog positions agree at the 99.7\% level\footnote{i.e.\ the Gaussian 
`3-$\sigma$' level, although as we used Rayleigh statistics we did not 
use 3-$\sigma$, but 99.7\%. This is smaller than the search 
radius used to merge distinct 1SXPS detections in to a unique source 
list, because the sky density of some external catalogs is high, and the 
number of spurious associations expected using a `5-$\sigma$' radius was 
unacceptably large}. SIMBAD\footnote{The SIMBAD and NED catalogs are 
dynamic entities: we cross-correlated against SIMBAD on 2013 June 10 and 
NED on 2013 September 6.} contains some sources from the 
facility-specific catalogs that we searched; such sources were only 
taken from the facility catalogs rather than repeating the match via 
SIMBAD/NED. We assumed zero position uncertainty for the SIMBAD,  
USNO-B1, 2MASS, NED and SDSS QSO catalogs, using just the 1SXPS position 
errors. For the USNO-B1, 2MASS and SDSS catalogs this is because 
their position errors are negligible compared to the 1SXPS errors. For 
SIMBAD and NED we are not able to specify the search radius as a 
function of source error in the remote query, and error information is 
not available in a uniform way; this may mean that for these catalogs 
the number of real matches which are not reported is higher than for the 
other catalogs. For the remaining catalogs we used the 1SXPS and 
catalog error added in quadrature. In some catalogs the systematic error 
is given only in the supporting documentation. This was added in 
quadrature to the catalog's statistical error when available. Details of 
the catalogs and their systematic errors are given in 
Table~\ref{tab:xcorr}, along with the number of 1SXPS sources which have 
a match in each catalog. Spatial coincidence alone of course 
does not guarantee association between the 1SXPS source and that in the 
external catalog. To estimate the number of spurious matches in this 
correlation, we shifted the position of each 1SXPS source at random by 
1--2\arcmin\ and repeated the correlation test. The number of matches 
found to these positions are also shown in Table~\ref{tab:xcorr}.

Due to the high sky density of the 2MASS and USNO-B1 catalogs, the number of expected spurious matches is very high at $>50\%$.
Indeed, there are frequently multiple matches from these catalogs to a single 1SXPS source, indicating that a 3-$\sigma$ spatial
coincidence in this case it a poor indicator of association.

We therefore ignored matches from these catalogs to estimate the number of new sources in our catalog: we 
found 68,638 1SXPS objects which are uncataloged (i.e.\ had no external catalog matches) in this case. However, as Table~\ref{tab:xcorr} shows, despite
these considerations there are 62,712 objects without a match in the USNO-B1 catalog and 99,353 without a match in the 2MASS source;
in total there are 20,390 sources in the 1SXPS catalog with no counterpart in any of the catalogs against which we performed a cross-correlation.

\begin{deluxetable*}{cccc}
\tablecaption{Catalogs cross-correlated with 1SXPS.}
\tablehead{
\colhead{Catalog}      &       \colhead{Systematic Error$^1$}  & \colhead{Number of matches$^2$}  &\colhead{Spurious matches$^3$} \\
} 
\startdata
SDSS Quasar Catalog DR5$^4$ &  & 1,781 & 9 (0.5\%) \\
XRTGRB$^5$ &  & 659 & 6 (1\%)\\
SwiftFT$^6$    &  &  9,154 & 268 (3\%) \\
1SWXRT$^7$     &  &  35,009 & 1,669 (5\%) \\
1CSC$^8$       &  &  6,334  & 340 (5\%) \\
3XMM DR4$^9$   &   &  19,649 & 1,381 (7\%)\\
ROSHRI$^10$     &  10\arcsec & 1,930  & 171 (9\%)\\
SIMBAD$^{11}$ & & 17,708 & 2,000 (11\%) \\
XMM SL1$^{12}$    &  17\arcsec  & 2,212 & 378 (17\%) \\
ROSPSPC$^{13}$    &  25\arcsec & 4,968 & 1,082 (22\%) \\
NED$^{14}$        &  & 49,098 & 14,761 (30\%) \\
USNO-B1$^{15}$ & & 88,812 & 48,718 (55\%) \\
2MASS$^{16}$ & & 52,171 & 33,549 (64\%) \\
\enddata
\tablecomments{$^1$ 90\%\ confidence \newline $^2$ Number of 1SXPS sources for which there is a counterpart
in the external catalog within 3-$\sigma$. \newline $^3$ The number of 1SXPS sources with a match
after the 1SXPS position has been moved by 1--2\arcmin; the value in brackets is this number as a percentage of the matches to 1SXPS positions
for the same external catalog.
\newline $^4$\cite{Schneider07}; $^5$ Taken from http://www.swift.ac.uk/xrt\_positions; see \cite{Evans09};
$^6$ \cite{puccetti11}; $^7$ \cite{delia13}; $^8$ \cite{iEvans10}; $^9$ http://xmmssc-www.star.le.ac.uk/Catalogue/3XMM-DR4/;
$^10$ http://heasarc.gsfc.nasa.gov/W3Browse/rosat/roshri.html
$^{11}$ http://simbad.u-strasbg.fr/simbad/ $^{12}$  \cite{saxton08}; $^{13}$ \cite{Voges99};
$^{14}$ http://ned.ipac.caltech.edu/; $^{15}$ \cite{Monet03}; $^{16}$ \cite{Skrutskie06}
}
\label{tab:xcorr}
\end{deluxetable*}

\subsection{Upper limit server}
\label{sec:upperlimit}

The 1SXPS website includes an upper limit server, which allows upper limits to be calculated
for any sky location covered by our catalog. If the location
was observed in more than one observation, upper limits can be calculated per observation,
or from the stacked image in which those observations are included. To calculate the upper limit
a 12-pixel radius circle is placed on the image at the location in question, and the number of events in
that circle is registered. The background level in this region is taken from the corresponding
background map. Then the Bayesian method of \cite{Kraft91} is used to determine the upper limit
on the source count-rate at the confidence level specified by the user. If the location requested matches that
of a \emph{Bad} detection which was discarded from the catalog (Section~\ref{sec:check}), this is also reported.

\section{Catalog characteristics and availability}
\label{sec:about}

The 1SXPS catalog contains 151,524 sources; 135,086 of which are not in flagged regions (Section~\ref{sec:visscreen}).
The median 90\%\ confidence  radial position error of the sources in the 
full catalog is 5.5\arcsec, including systematic errors, and the median 
0.3--10 keV flux is 3\tim{-14} erg \cms\ s$^{-1}$. The total exposure 
time of the observations in the catalog is 147 Ms, spread over 1905 square degrees on the sky.
10\%\ of the exposure time lies at a Galactic latitude $|b|<3\deg$; 14\%\ of 1SXPS sources lie in this 
latitude range, showing as expected an overdensity of sources in the Galactic plane compared
to the sky as a whole.

The catalog of sources and their properties is available for download as 
a FITS or ASCII table from the 1SXPS website: 
\hbox{http://www.swift.ac.uk/1SXPS}. Table~\ref{tab:srcdef} describes 
the columns in the catalog. This website also provides simple and 
comprehensive search facilities, a detailed web page for each source and 
each dataset, as well as the upper limit server 
(Section~\ref{sec:upperlimit}). The main catalog file is also available 
through Vizier (catalog ID: IX/43). 

A table of external catalog cross correlations (Section~\ref{sec:xcorr}) 
is available from the site above, as are tables giving information about 
the individual detections and the datasets. These tables are described 
in Tables~\ref{tab:fielddef}--~\ref{tab:xcorrdef}. We request that 
publications which make use of this catalog state in the 
acknowledgements: \emph{This work made use of data supplied by the UK 
Swift Science Data Centre at the University of Leicester.} as well as 
citing this paper.

When selecting objects from the tables, the combination of detection 
flags and field flags gives great control over whether sensitivity or 
purity is prioritized. The catalog website also provides postage-stamp 
images of each source and images of each dataset; when considering 
sources with detection flags $\ge8$ it is recommended to view these 
images to help judge their reliability. For the rest of 
this paper we conservatively defined a `clean' subsample of the catalog, 
comprising all objects with detection and field flags both $<2$ (i.e.\  
\good\ or \reasonable, and from a field that is either OK, or affected 
only by artifacts but not in the region covered by the artifact): there are 98,762 such sources in the catalog.

{
\LongTables

\begin{deluxetable*}{llll}
\tablecaption{Contents of the main catalog table (`sources')}
 \tablehead{
 \colhead{Field}     & Units   &       \colhead{Description} & \colhead{Has errors?$^1$} \\
}
\startdata
\multicolumn{4}{c}{\emph{Name and position}} \\
Name &  & Unique identifier, of the form: 1SXPS JHHMMSS.S+DDMMSS & \\
RA & degrees  & Right Ascension  (J2000)  & \\
Decl & degrees & Declination (J2000) & \\
Err90 & arcsec & 90\%\ conf. radial position error &  \\
AstromType & & The provenance of the astrometry used for the source position. \\
& & 0=\swift\ star tracker, 1=XRT/2MASS correlation \\
l & degrees & Galactic longitude \\
b & degrees & Galactic latitude \\
OffAxis & arcmin & The mean off-axis angle of this source from the \\
& & observations in which it was detected \\
\multicolumn{4}{c}{\emph{Exposure details}} \\
Exposure & s & The total exposure at the source location \\
StartDate & UT & The calendar date of the start of the first observation \\
& &  of the location of this source \\
StopDate & UT & The calendar date of the end of the last observation \\
& &  of the location of this source \\
NumObs & & The number of observations of the location of the source \\
NumDetObs & & The number of observations in which the source was detected \\
\multicolumn{4}{c}{\emph{Flag details}} \\
DetFlag & & The best detection flag from all detections of this source \\
Fieldflag & & The best field flag  from all detections of this source \\
DetFlag\_band[0--4] & & The best detection flag in each band, from all \\
& &  detections of the source in that band \\
\multicolumn{4}{c}{\emph{Count-rate and variability information}} \\
Rate\_band[0--4] & ct s$^{-1}$ & The mean count-rate of the source in each band & yes \\
Counts\_band[0--4] & & The number of counts measured in the region of \\
& &  the source in each band \\
BGCounts\_band[0--4] & & The number of counts in the background map \\
& &  in the region of the source in each band\\
CF\_band[0--4] & & The count-rate correction factor ($\kappa$)  for the \\
& & source in each band \\
PvarPchiSnapshot\_band[0--4] & & The probability that the source is constant \\
& &  between snapshots in band 0--4,\\
& &  deduced via the Pearson's \chisq\ test \\
PvarPchiObsID\_band[0--4] & & The probability that the source is constant \\
& &  between observations in band 0--4,\\
& &  deduced via the Pearson's \chisq\ test \\
HR1 & & The HR1 hardness ratio & yes \\
HR2 & & The HR2 hardness ratio & yes \\
\multicolumn{4}{c}{\emph{Flux and spectral information}} \\
GalNH & \cms & The Galactic absorption column density in the \\
& &  direction of the source \\
whichPow & & The provenance of the summary spectral fields for \\
& &  the power-law model.\\
& &  0=fixed spectrum, 1=HR-derived, 2=fitted spectrum \\
whichAPEC & & The provenance of the summary spectral fields for \\
& &  the APEC model.\\
& &  0=fixed spectrum, 1=HR-derived, 2=fitted spectrum \\
\multicolumn{4}{l}{\emph{Summary spectral information$^2$}} \\
PowECFO & erg \cms\ ct $^{-1}$ & The counts-to-observed-flux energy conversion factor \\
& & derived from the power-law spectrum \\
PowECFU & erg \cms\ ct $^{-1}$ & The counts-to-unabsorbed-flux energy conversion factor \\
& & derived from the power-law spectrum \\
PowFlux & erg \cms\ s$^{-1}$ & The mean observed source flux derived \\
& &  from the power-law spectrum & yes \\
PowPeakFlux & erg \cms\ s$^{-1}$ & The peak$^3$ observed source flux derived from the \\
& &   power-law spectrum & yes \\
PowUnabsFlux & erg \cms\ s$^{-1}$ & The mean unabsorbed source flux derived from the \\
& &  power-law spectrum & yes \\
PowPeakUnabsFlux & erg \cms\ s$^{-1}$ & The peak$^3$ unabsorbed source flux derived from the\\
& &  power-law spectrum & yes \\
APECECFO & erg \cms\ ct $^{-1}$ & The counts-to-observed-flux energy conversion factor \\
& & derived from the APEC spectrum \\
APECECFU & erg \cms\ ct $^{-1}$ & The counts-to-unabsorbed-flux energy conversion factor  \\
& & derived from the APEC spectrum \\
APECFlux & erg \cms\ s$^{-1}$ & The mean observed source flux derived \\
& & from the APEC spectrum & yes \\
APECPeakFlux & erg \cms\ s$^{-1}$ & The peak$^3$ observed source flux derived from the\\
& &   APEC spectrum & yes \\
APECUnabsFlux & erg \cms\ s$^{-1}$ & The mean unabsorbed source flux derived from the\\
& &   APEC spectrum & yes \\
APECPeakUnabsFlux & erg \cms\ s$^{-1}$ & The peak$^3$ unabsorbed source flux derived from the\\
& &   power-law spectrum & yes \\
\multicolumn{4}{l}{\emph{Detailed spectral information}} \\
FixedPowECFO & erg \cms\ ct $^{-1}$ & The counts-to-observed-flux energy conversion factor \\
& & derived from the fixed power-law spectrum \\
FixedPowECFU & erg \cms\ ct $^{-1}$ & The counts-to-unabsorbed-flux energy conversion factor  \\
& & derived from the fixed power-law spectrum \\
FixedPowFlux & erg \cms\ s$^{-1}$ & The mean observed source flux derived \\
& &  from the fixed power-law spectrum & yes \\
FixedPowUnabsFlux & erg \cms\ s$^{-1}$ & The mean unabsorbed source flux derived \\
& &  from the fixed power-law spectrum & yes \\
FixedAPECECFO & erg \cms\ ct $^{-1}$ & The counts-to-observed-flux energy conversion factor \\
& &  derived from the fixed APEC spectrum \\
FixedFixed APECECFU & erg \cms\ ct $^{-1}$ & The counts-to-unabsorbed-flux energy conversion factor \\
& & derived from the fixed APEC spectrum \\
FixedAPECFlux & erg \cms\ s$^{-1}$ & The mean observed source flux derived \\
& &  from the fixed APEC spectrum & yes \\
FixedAPECUnabsFlux & erg \cms\ s$^{-1}$ & The mean unabsorbed source flux derived \\
& &  from the fixed APEC spectrum & yes \\
InterpPowECFO & erg \cms\ ct $^{-1}$ & The counts-to-observed-flux energy conversion factor \\
& & derived from the HR-derived power-law spectrum \\
InterpPowECFU & erg \cms\ ct $^{-1}$ & The counts-to-unabsorbed-flux energy conversion factor \\
& & derived from the HR-derived power-law spectrum \\
InterpPowFlux & erg \cms\ s$^{-1}$ & The mean observed source flux derived from the HR-derived \\
& &  power-law spectrum & yes \\
InterpPowUnabsFlux & erg \cms\ s$^{-1}$ & The mean unabsorbed source flux derived \\
& &  from the HR-derived power-law spectrum & yes \\
InterpPowNH & \cms\  & The absorption column density derived from the \\
& &  HR-derived power-law spectrum & yes \\
InterpPowGamma & erg  & The power-law photon index derived from the \\
& &  HR-derived power-law spectrum & yes \\
InterpAPECECFO & erg \cms\ ct $^{-1}$ & The counts-to-observed-flux energy conversion \\
& &  factor derived from the HR-derived APEC spectrum \\
InterpAPECECFU & erg \cms\ ct $^{-1}$ & The counts-to-unabsorbed-flux energy conversion \\
& &  factor derived from the HR-derived APEC spectrum \\
InterpAPECFlux & erg \cms\ s$^{-1}$ & The mean observed source flux derived \\
& &  from the HR-derived APEC spectrum & yes \\
InterpAPECUnabsFlux & erg \cms\ s$^{-1}$ & The mean unabsorbed source flux derived \\
& &  from the HR-derived APEC spectrum & yes \\
InterpAPECNH & \cms\  & The absorption column density derived \\
& &  from the HR-derived APEC spectrum & yes \\
InterpAPECkT & keV  & The plasma temperature derived from \\
& &  the HR-derived APEC spectrum & yes \\
P\_pow & & For sources without an HR-derived value, the probability of \\
& &  measuring the (HR1,HR2) value of this source \\
& &  if it had an power-law spectrum \\
P\_APEC & & For sources without an HR-derived value, the probability \\
& &  of measuring the (HR1,HR2) value of this sourc \\
& & e if it had an APEC spectrum \\
FittedPowECFO & erg \cms\ ct $^{-1}$ & The counts-to-observed-flux energy conversion factor \\
& &  derivedfrom the fitted power-law spectrum \\
FittedPowECFU & erg \cms\ ct $^{-1}$ & The counts-to-unabsorbed-flux energy conversion factor \\
& &  derived from the fitted power-law spectrum \\
FittedPowFlux & erg \cms\ s$^{-1}$ & The mean observed source flux derived \\
& &  from the fitted power-law spectrum & yes \\
FittedPowUnabsFlux & erg \cms\ s$^{-1}$ & The mean unabsorbed source flux derived \\
& &  from the fitted power-law spectrum & yes \\
FittedPowNH & \cms\  & The absorption column density derived \\
& &  from the fitted power-law spectrum & yes \\
FittedPowGamma & erg  & The power-law photon index derived \\
& &  from the fitted power-law spectrum & yes \\
FittedPowChi & & \chisq\ of the power-law spectral fit \\
FittedPowDOF & & Degrees of freedom in the power-law spectral fit \\
FittedPowRedChi & & \rchisq\ in the power-law spectral fit \\
FittedAPECECFO & erg \cms\ ct $^{-1}$ & The counts-to-observed-flux energy conversion factor  \\
& &  derived from the fitted APEC spectrum \\
FittedAPECECFU & erg \cms\ ct $^{-1}$ & The counts-to-unabsorbed-flux energy conversion factor  \\
& & derived from the fitted APEC spectrum \\
FittedAPECFlux & erg \cms\ s$^{-1}$ & The mean observed source flux derived \\
& &  from the fitted APEC spectrum & yes \\
FittedAPECUnabsFlux & erg \cms\ s$^{-1}$ & The mean unabsorbed source flux derived \\
& &  from the fitted APEC spectrum & yes \\
FittedAPECNH & \cms\  & The absorption column density derived \\
& &  from the fitted APEC spectrum & yes \\
FittedAPECkT & keV  & The plasma temperature derived from the \\
& & fitted APEC spectrum & yes \\
FittedAPECChi & & \chisq\ of the APEC spectral fit \\
FittedAPECDOF & & Degrees of freedom in the APEC spectral fit \\
FittedAPECRedChi & & \rchisq\ in the APEC spectral fit \\
\multicolumn{4}{l}{\emph{Cross-correlation information$^2$}} \\
Numxcorr & & The number of matches in the external catalogs \\
Numxcorr\_slim & & The number of matches in the external catalogs, \\
& &  excluding USNO-B1 and 2MASS \\
isROSHRI  &  & Whether the object does (1) or does not (0) match \\
& &  an object in the Rosat HRI catalog \\
isROSPSPC  &  & Whether the object does (1) or does not (0) match \\
& &  an object in the Rosat PSPC catalog \\
is3XMM  &  & Whether the object does (1) or does not (0) match \\
& &  an object in the 3XMM DR4 catalog \\
isXMMSL1  &  & Whether the object does (1) or does not (0) match \\
& &  an object in the XMMSL1 XMM-Newton Slew Survey \\
isSwiftFT  &  & Whether the object does (1) or does not (0) match \\
& &  an object in the Swift-FT catalog \\
is1SWXRT  &  & Whether the object does (1) or does not (0) match \\
& &  an object in the 1SWXRT catalog \\
isXRTGRB  &  & Whether the object does (1) or does not (0) match \\
& &  a cataloged XRT position of a Gamma Ray Burst \\
isSDSSQSO  &  & Whether the object does (1) or does not (0) match \\
& &  an object in the SDSS QSO DR 5 catalog \\
is2MASS  &  & Whether the object does (1) or does not (0) match \\
& &  a 2MASS source \\
isUSNOB1  &  & Whether the object does (1) or does not (0) match \\
& &  a USNO-B1 source \\
isSIMBAD  &  &Whether the object does (1) or does not (0) match \\
& &  a SIMBAD object \\
xcorrIDs & & A semi-colon delimited list of the identifiers \\
& & of the matched sources \\
\enddata
\label{tab:srcdef}
\tablecomments{$^1$ This is `no' unless stated. For a field with errors, there are two error fields, \emph{fieldname}\_pos and 
\emph{fieldname}\_neg. $^2$This is taken from the detailed spectral information, for the method given in the \emph{whichPow} and \emph{whichAPEC}
fields. $^3$ The peak flux is derived using the summary ECF and the count-rate in of the brightest bin in the total band per-snapshot light curve.
}

\end{deluxetable*}

\clearpage

\begin{deluxetable*}{lll}
\tablecaption{Contents of the `Datasets' catalog table}
 \tablehead{
 \colhead{Field}     & Units   &       \colhead{Description}  \\
}
\startdata
ID & & The unique identifier of the dataset. For observations \\
& &  this is the 11-digit ObsID. For stacked images \\
& &  it is the number of the image. \\
RA & degrees  & Right Ascension of the field center (J2000) \\
Decl & degrees & Declination of the field center (J2000) \\
l & degrees & The Galactic longitude of the field center \\
b & degrees & The Galactic latitude of the field center \\
IsStacked & & Indicates whether this is a stacked image (1) or not (0) \\
Exposure & s & The exposure time in the dataset \\
FieldBG\_band[0--4] & ct s$^{-1}$ pixel$^{-1}$ & The mean background level in each band. \\
Numsrc\_band[0--4] & & The number of sources in this image in each band.\\
NumOK\_band[0--4] & & The number of \good\ and \reasonable\ sources in each band.\\
MedianNNDist\_band[0--4] & & The median distance between the sources in each band's image. \\
Date\_start & UT & The calendar date of the observation start \\
Date\_stop & UT & The calendar date of the observation end \\
FieldFlag & & The field flag \\
NumSnapshots & & The number of snapshots in the dataset. \\
AstromErr & arcsec & The 90\%\ confidence uncertainty in the astrometric solution \\
& &  for this field derived using 2MASS (Section~\ref{sec:xastrom}). \\
StackedImage & & For observations: the ID of the stacked image in which this observation is included.\\
             & & For stacked images: the IDs of any stacked images \\
& &  which overlap this one. \\
\enddata
\label{tab:fielddef}
\end{deluxetable*}
\clearpage

\begin{deluxetable*}{llll}
\tablecaption{Contents of the `Detections' catalog table}
 \tablehead{
 \colhead{Field}     & Units   &       \colhead{Description} & \colhead{Has errors?} \\
}
\startdata
DetID & & A unique identifier for this detection \\
ObsID & & The unique 11-digit obsID of the dataset the detection occurred in. \\
Band & & The band in which the detection occured, \\
& & (0=total, 1=soft, 2=medium, 3=hard) \\
DetFlag & & The detection flag as an integer value  \\
img\_x$^1$ & pixels & The x-location of the detection in XRT {\sc sky} coordinates \\
img\_y$^1$ & pixels & The y-location of the detection in XRT {\sc sky} coordinates \\
OffAxis & arcmin & The mean off-axis angle of the detection in this observation \\
RA & degrees & RA (J2000) of the detection using the star tracker astrometry  & yes (statistical only) \\
Dec & degrees & Declination (J2000) of the detection using the star tracker astrometry& yes (statistical only) \\
Err90 & arcsec & 90\%\ conf. radial position error, statistical+systematic \\
RA\_corr & degrees & RA (J2000) of the detection using 2MASS/XRT astrometry &  \\
Dec\_corr & degrees & Declination (J2000) of the detection using 2MASS/XRT astrometry & \\
Err90\_corr & arcsec & 90\%\ conf. radial position error using 2MASS/XRT astrometry \\
l & degrees & Galactic longitude of the detection \\
b & degrees & Galactic latitude of the detection \\
Counts &  & Number of events in the count-rate extraction region  \\
BGCts &  & The expected number of background events in the above region \\
FieldExposure & s & The on-axis exposure of the dataset the detection is in. \\
CF & & The count-rate correction factor ($\kappa$)  \\
Rate & ct s$^{-1}$ & The count rate of the detection. & yes \\
ExposureFraction &  & The exposure time at the location of the \\
& & detection divided by the on-axis exposure \\
Cstat & & The \cstat\ value from the PSF fit \\
Cstat\_nosrc & & The \cstat\ value calculated with normalization=0 \\
LogLikelihood &  & The log-likelihood of the detection. \\
SNR & & The SNR of the detection \\
Celldet\_width & pixels & The size of the cell in which the detection was made. \\
PSF\_Radius & pixels & The radius of the circular region used in PSF fitting. \\
PSF &  & Which PSF profile was selected by PSF fitting. \\
ol\_warn & magnitude &The number of magnitudes brighter than the \\
& &  warning level of any cataloged star within \\
& &  30\arcsec\ of the detection. \\
FieldFlag & & The flag associated with the dataset the detection is in. \\
NNDist & arcsec & The distance to the nearest other detection in this image \\
OKNNDist & arcsec & The distance to the nearest \good\ or \reasonable\ detection in this image. \\
Num\_snapshots & & How many snapshots are in the image containing the detection. \\
ImageBG & ct s$^{-1}$ pixel$^{-1}$ & The mean background level in the image, \\
& &  according to the background map.\\
MergeRadius & pixels & The radius over which other detections in this image \\
& &  are assumed to be aliases of this detection. \\
SourceID & & The identifier of the unique 1SXPS source this to \\
& &  which this detection corresponds. \\
\enddata
\label{tab:detdef}
\tablecomments{$^1$ The {\sc sky} coordinate system for an image depends on the position information used 
process the raw XRT data, thus may not be the same for user-processed data.}
\end{deluxetable*}
\clearpage

\begin{deluxetable*}{lll}
\tablecaption{Contents of the `Cross Correlations' catalog table}
 \tablehead{
 \colhead{Field}     & Units   &       \colhead{Description}  \\
}
\startdata
1SXPS\_ID & & The name of the 1SXPS source \\
ExtCat\_ID & & The name of the source in the external catalog \\
Catalog & & The catalog containing the matched source \\
Distance & arcsec & The distance between the 1SXPS source and external catalog source \\
RA & degrees & The RA (J2000) of the source in the external catalog \\
Decl & degrees & The Declination (J2000) of the source in the external catalog \\
Err90 & arcsec & The 90\%\ confidence radial uncertainty in the external catalog position, including any systematic \\
\enddata
\label{tab:xcorrdef}
\end{deluxetable*}

}

\section{Verification}
\label{sec:verify}

We used simulations to verify the accuracy of the catalog, making these as realistic as possible
by basing our simulations on real data. To do this we identified XRT observations of 2XMMiDR3 \citep{Watson09} fields,
selected from that catalog all sources expected to contribute at least two events to the XRT image
(assuming a typical AGN spectrum: $N_H=3\tim{20}$ \cms, $\Gamma=1.7$), and visually inspected the XRT image to
ensure that this list identified all objects in the field. We then passed this source list to our background
map software, which created a model of the background in the real XRT image. This model then forms the basis 
of the simulations. We did this for a range of different positions on the sky and XRT exposure times.

To simulate an image we then used the background map just created, with the 
corresponding exposure map to measure the number of background counts, 
$\mu_i$, in each pixel $i$. For each pixel in the image we drew the 
number of events to simulate at random from a Poisson distribution with 
a mean of $\mu_i$. To add sources to the image we randomly drew from the 
$\log N-\log S$ distribution of extragalactic sources from 
\cite{Mateos08}. For each source we randomized the position on the CCD, 
and then simulated $C$ events, where $C$ was drawn randomly from a 
Poisson distribution with a mean equal to the number of events expected 
from that source on-axis. These events were folded through the 
instrumental PSF to locate the specific pixel in which the photon fell. 
If the exposure map value at this pixel was less than the on-axis 
exposure value, a random number between 0 and 1 was generated. If this 
number was less than the fractional exposure of the pixel in question, 
the photon was added to the image, otherwise it was discarded. In this 
way we build up a realistic XRT image.

Although we had a discrete set of `seed' images from which we could simulate data, by selectively excluding snapshots 
from those images, we were able to simulate a larger selection of exposure times than would be given simply by considering the 
seed images as unit elements. Similarly, we could simulate a range of background levels by multiplying the seed background map by an appropriate
value. This allowed us to test our catalog software on a range of exposure times and background levels which mirrors that of the 
data in the catalog.

\subsection{Background maps}
\label{sec:verifybg}

\begin{figure}
\begin{center}
\includegraphics[height=8.1cm,angle=-90]{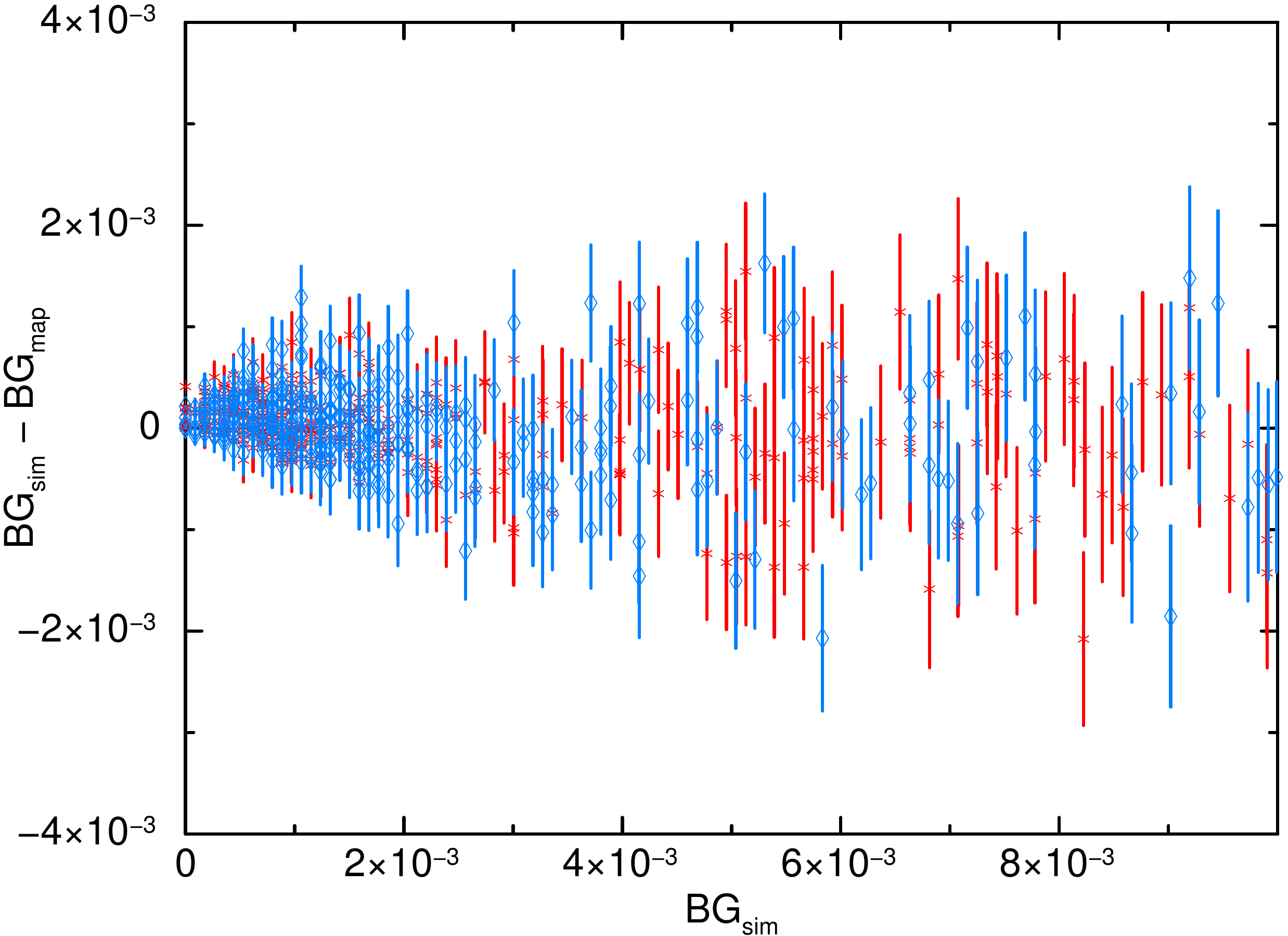}
\caption{Comparison of the background measured directly from the simulated image (BG$_{\rm sim}$) with that
measured from a background map (BG$_{\rm map}$) constructed from the simulated image.
The red stars show the simulations with no sources included, the blue diamonds the simulations containing
a source.
}
\label{fig:bgtest}
\end{center}
\end{figure}

To confirm that our background mapping was working correctly we simulated 400 images, with the background level and 
exposure time drawn at random from the distribution of those values seen in the catalog. Since these 
contain no sources, the true background level of each image can be measured directly.
We then used our software to build a background map of these images and measured the background level from these maps,
to compare with the true value.  We measured the background by placing a circle of radius 60 pixels
at a random location on the image and
taking the mean value of all pixels in this circle with non-zero exposure. The same circle was
used for an image and the corresponding
background map, but a different circle was randomly placed for each simulation. The 60-pixel radius is much larger
than the source extraction region used in the catalog, but is needed to reduce the 
magnitude of the Poisson uncertainty on the measurement of the simulated image. Fig.~\ref{fig:bgtest} shows the results of these
tests, confirming that the background mapping tool performs well. 

We performed a further 400 simulations independent of the set used 
above. This time a single source was added to the simulated image, 
although we also saved the source-less image, from which we measured the 
true background level. We then ran our source detection code on the 
image including the source. This detected the source and built a map of 
the underlying background. As  Fig.~\ref{fig:bgtest} shows, the 
reconstructed background in these cases still accurately reflects the 
true value: a \chisq\ test for the model BG$_{\rm sim}-$BG$_{\rm map}=0$
applied to these data gives \rchisq=0.84, for 788 degrees of freedom.


\subsection{Count-rate reconstruction}
\label{sec:verifyrate}

\begin{figure}
\begin{center}
\includegraphics[width=8.1cm]{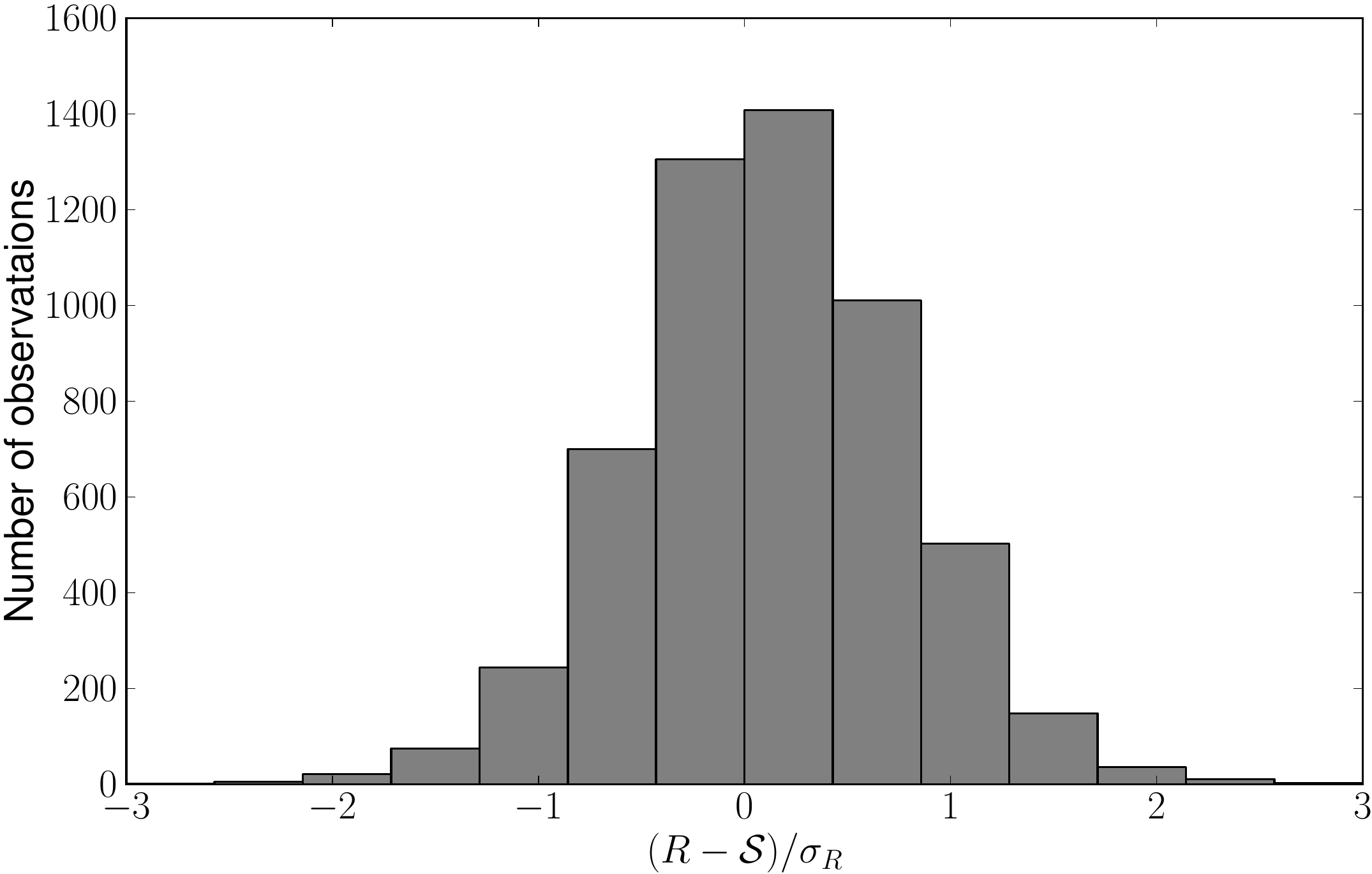}
\includegraphics[width=8.1cm]{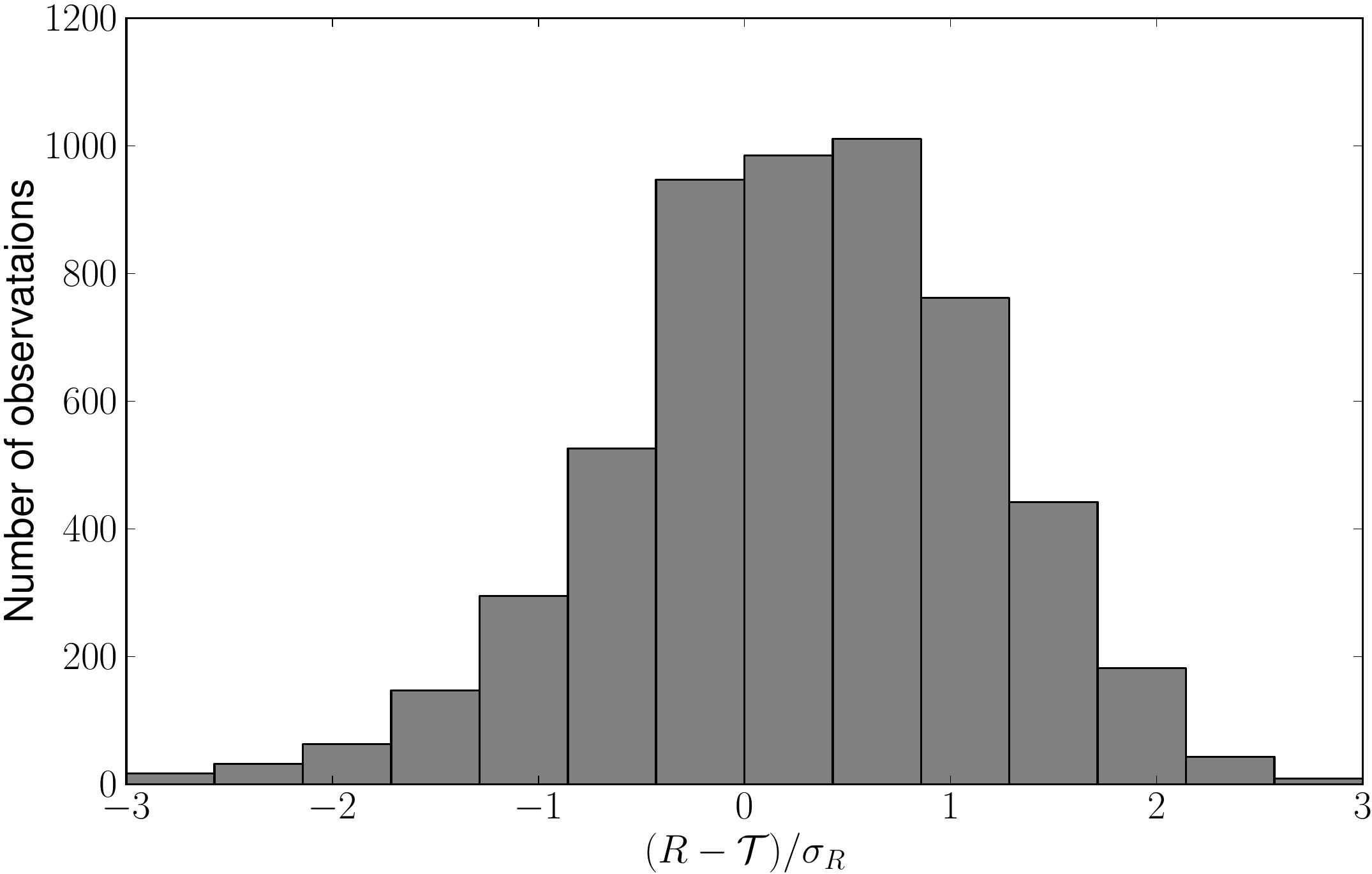}
\caption{Histogram showing the accuracy of our count-rate reconstruction, based on simulation.
\emph{Top:} The difference between the measured and simulated source count rate, divided by the error on the measured value.
\emph{Bottom:} The difference between the measured and true source count rate, divided by the error on the measured value. The asymmetry in this plot
is due to the Eddington bias.
}
\label{fig:verifyrate}
\end{center}
\end{figure}

To test whether the source count-rate was adequately reconstructed, we performed a further 5,000 simulations,
this time with multiple sources per image, as described in Section ~\ref{sec:verify}.
For each source we drew the flux from the $\log N-\log S$ distribution, multiplied it by the image exposure time
and folded it though a typical AGN spectrum to obtain the expected number of XRT events, $\mathcal{T}_c$. To incorporate
Poisson processes we then drew a number $\mathcal{S}_c$ from a Poisson distribution with a mean of $\mathcal{T}_c$; this 
($\mathcal{S}_c$) was the number of events which were actually put into the simulation. These events
are folded through the PSF and exposure map (Section ~\ref{sec:verify}); the number which are actually included in the
simulated image is $A_c$. Each of these numbers ($\mathcal{T}_c, \mathcal{S}_c, A_c$) can be converted to a count-rate ($\mathcal{T}, \mathcal{S}, A$) by dividing by the
on-axis exposure time of the simulated image.

We ran the catalog software on these 5,000 simulated images to detect and characterize the sources, and then
compared the count-rates thus obtained with the simulated count-rates.
The top panel of Fig.~\ref{fig:verifyrate} shows the distribution of $(R-\mathcal{S})/\sigma_R$, where $R$ and $\sigma_R$ are
the source count-rate and error returned by the catalog software. This shows that our software is accurately reconstructing
the count rates. The non-zero width of the distribution arises because of the PSF corrections and Poisson noise: if a source
is located on the detector such that, on average, 30\% of the simulated events are lost (i.e. $A/\mathcal{S}=0.7$) then 
the catalog software (correctly) applies a correction of $\kappa=1/0.7$ to the measured count-rate. However due to Poisson
processes, the values of $A/\mathcal{S}$ in the simulations show scatter around this mean value. Fig.~\ref{fig:verifyrate}
shows that this scatter is relatively narrow (a Gaussian fit has $\sigma\til 0.6$), and adding it to the count-rate
uncertainty makes negligible difference to that value. Thus this effect can be safely neglected.

The bottom panel of Fig.~\ref{fig:verifyrate} shows the distribution of $(R-\mathcal{T})/\sigma_R$. As can be seen, this
distribution is significantly skewed with the catalog tending to overestimate the true count-rate. This is 
simply the result of the Eddington bias \citep{Eddington40}: if the true source count-rate is close to the
detector limit then we detect those sources which Poisson noise makes appear brighter, but not those which are made
fainter.

\subsubsection{Eddington bias}
\begin{figure*}
\begin{center}
\includegraphics[width=16.2cm]{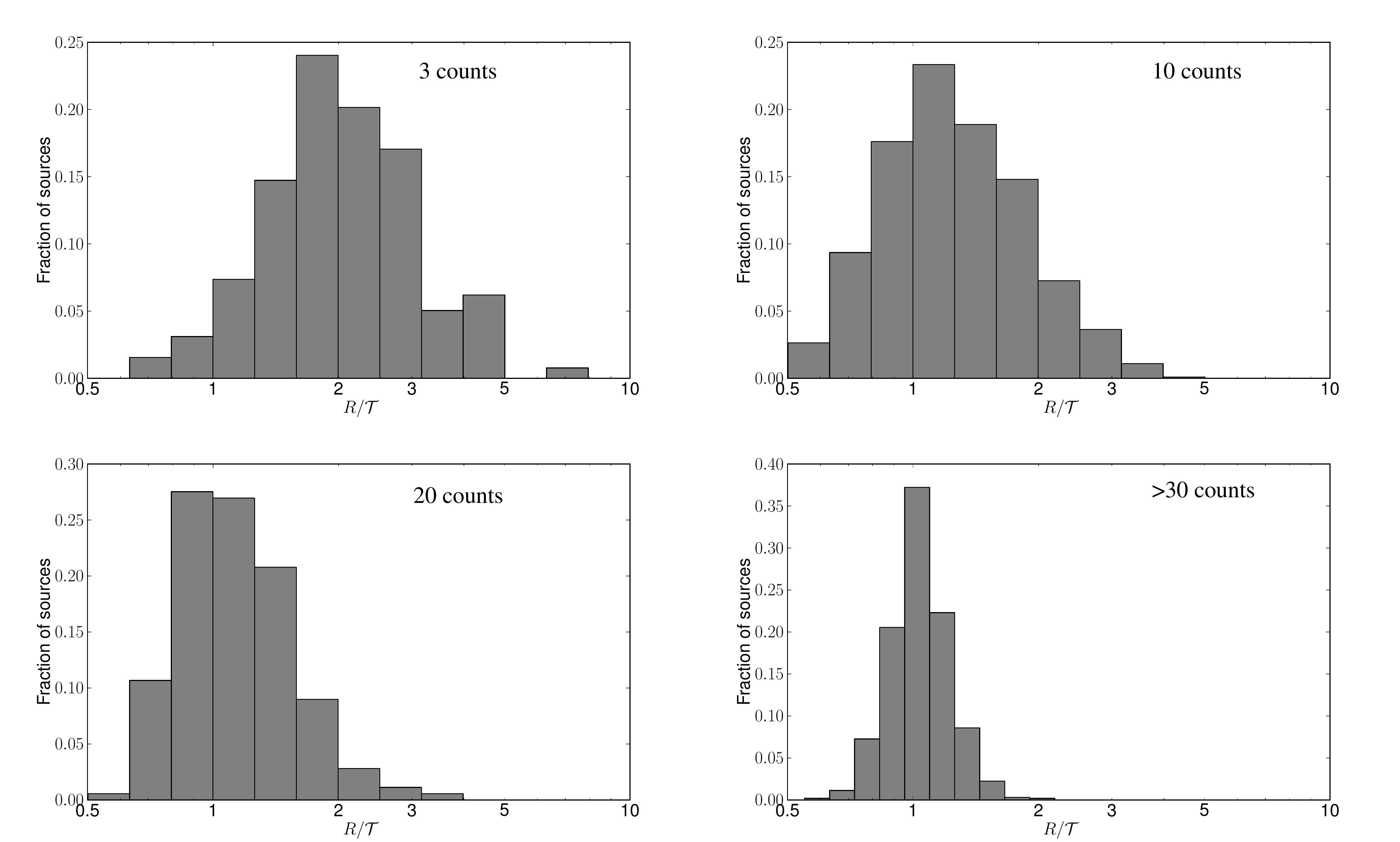}
\caption{The effect of the Eddington Bias, showing the ratio of the measured count-rate to the true count-rate ($R/\mathcal{T}$)
as a function of the number of simulated counts, $A_c$.}
\label{fig:EddBias}
\end{center}
\end{figure*}

To explore the magnitude of the Eddington Bias in our data, we simulated a further 20,000 images, again with the exposure time
and background level drawn at random from the distributions seen in the catalog, and with the source fluxes drawn from a $\log N-\log S$
distribution. We then ran our catalog software on those images, recording both the `true' count-rate from the simulation ($\mathcal{T}$)
and the count-rate $R$ determined by our software. In Fig.~\ref{fig:EddBias} we show the distribution of the ratio $R/\mathcal{T}$ 
as a function of how many simulated events there were ($A_c$) for the source in question. This shows that (unsurprisingly) the Eddington bias
is very strong for the faintest sources in the catalog, with the count-rates determined typically a factor of 2 too high. Although
this bias lessens as we move to brighter sources, the distribution of rates recovered is still significantly asymmetric at 
$A_c$=20, however for sources with at least 30 events, the effect of the Eddington bias has all but disappeared.

\subsection{Variability test}
\label{sec:verifyvar}

\begin{figure}
\begin{center}
\includegraphics[width=8.1cm]{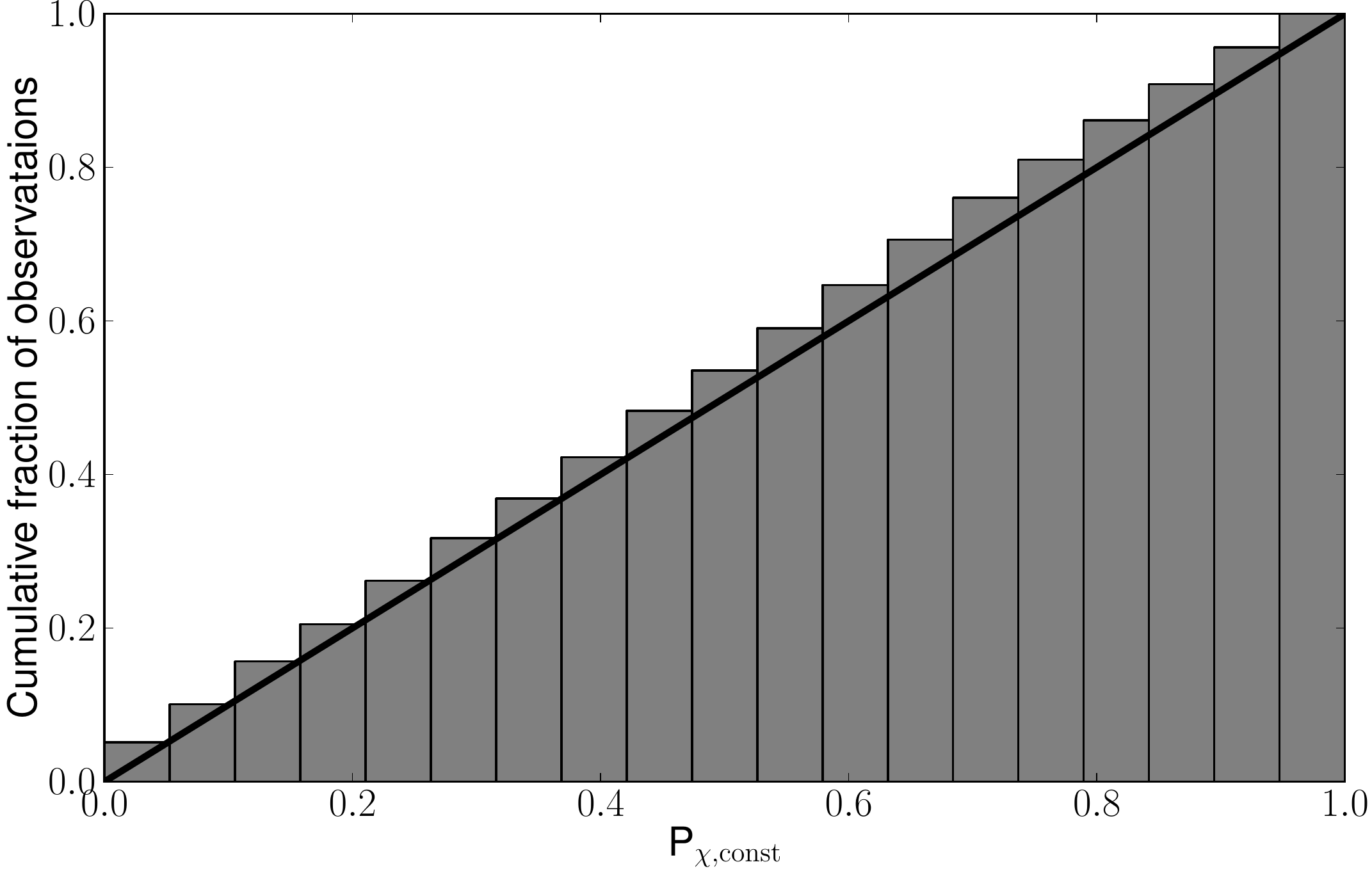}
\caption{The cumulative probability distributions from the Pearson's \chisq\ variability test applied to the 
constant sources in 5,000 simulated images. The black line shows the expected result which is well matched by the data.
}
\label{fig:verifyvar}
\end{center}
\end{figure}

We performed the Pearson's \chisq\ tests for variability on the sources in the 5,000 simulations
created for Section~\ref{sec:verifyrate}. Since these sources are simulated with constant intensity (which is the null
hypothesis of these tests) we expect that 10\%\ of the sources will have a $P<0.1$ etc.
Fig.~\ref{fig:verifyvar} shows that this is the result obtained. This does not provide information
on how strong variability has to be before it is detected, however this is a function of variability type,
exposure, source brightness, light curve sampling etc. and should be determined on a per-source basis.

\subsection{Spectroscopy}
\label{sec:verifyspec}

\begin{figure}
\begin{center}
\includegraphics[height=8.1cm,angle=-90]{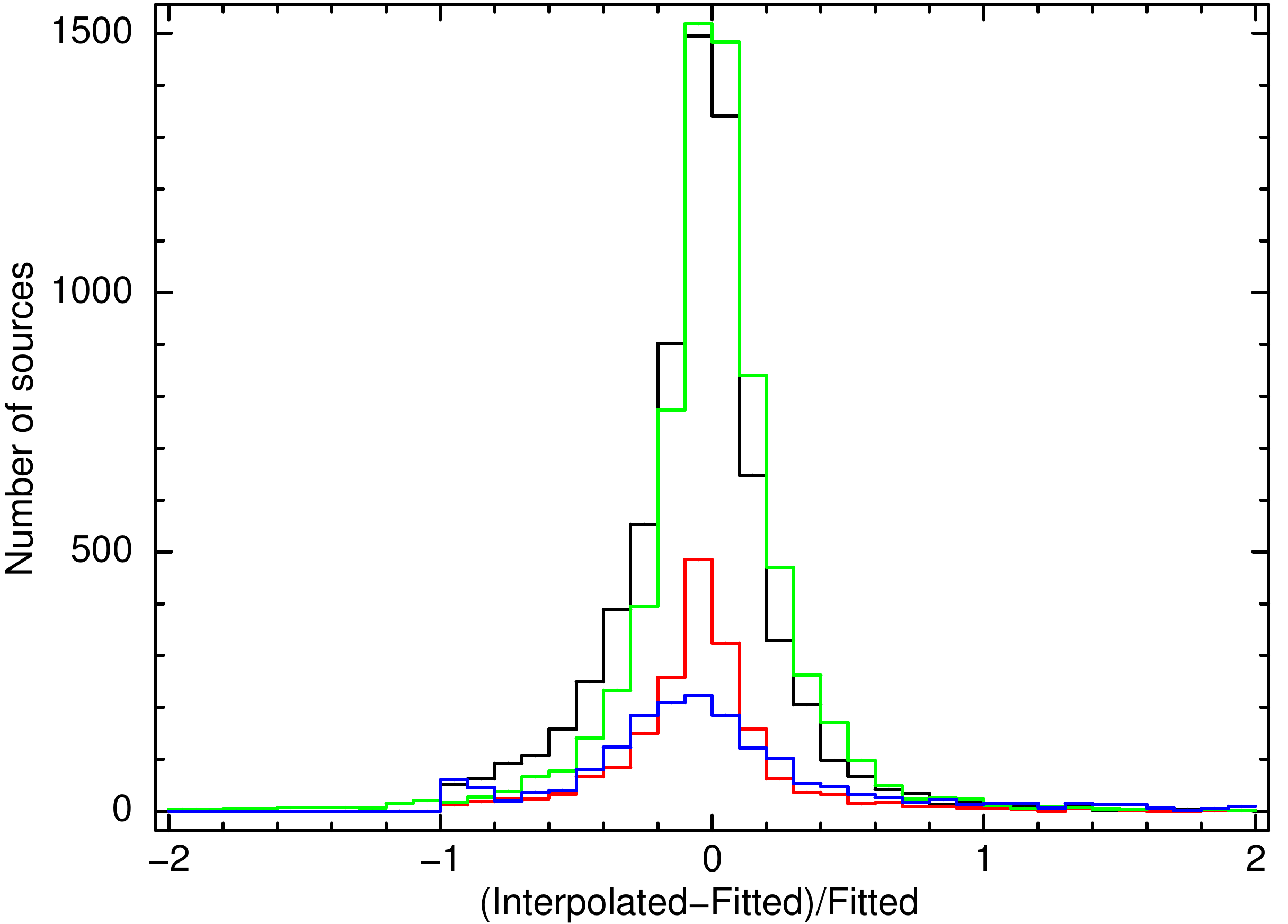}
\caption{The difference between the spectral parameters derived from the hardness ratio and those
from the spectral fit, divided by the spectrally-fitted value. Only sources where the spectral fit had $\rchisq<1.5$ are shown.
\emph{Black and red}: Observed flux from a power-law and APEC spectrum respectively.
\emph{Green}: Photon index from a power-law spectrum.
\emph{Blue}: Plasma temperature from an APEC spectrum.
}
\label{fig:verifyspec}
\end{center}
\end{figure}

The distribution of \rchisq\ from the power-law and APEC model spectral fits shows a clustering around \rchisq=1 for both spectral models, as expected
if those models are good representations of the data. About 25\%\ of fits have $\rchisq\gg1$, these represent cases where the simple spectral models
we have used are not appropriate and more complex (e.g.\ multi-temperature) emission processes are likely involved.
For those sources for which we have both a spectral fit with $\rchisq<1.5$ and an estimate of the spectral parameters 
derived from the hardness ratios, we show in Fig.~\ref{fig:verifyspec} a histogram of the HR$-$Fit/Fit, for both the observed
flux and the emission parameter. This shows that the spectral parameters derived from the hardness ratios are reasonable.

\section{Quality flags, false positive rate and catalog completeness}
\label{sec:falsep}

\begin{figure}
\begin{center}
\includegraphics[height=8.1cm,angle=-90]{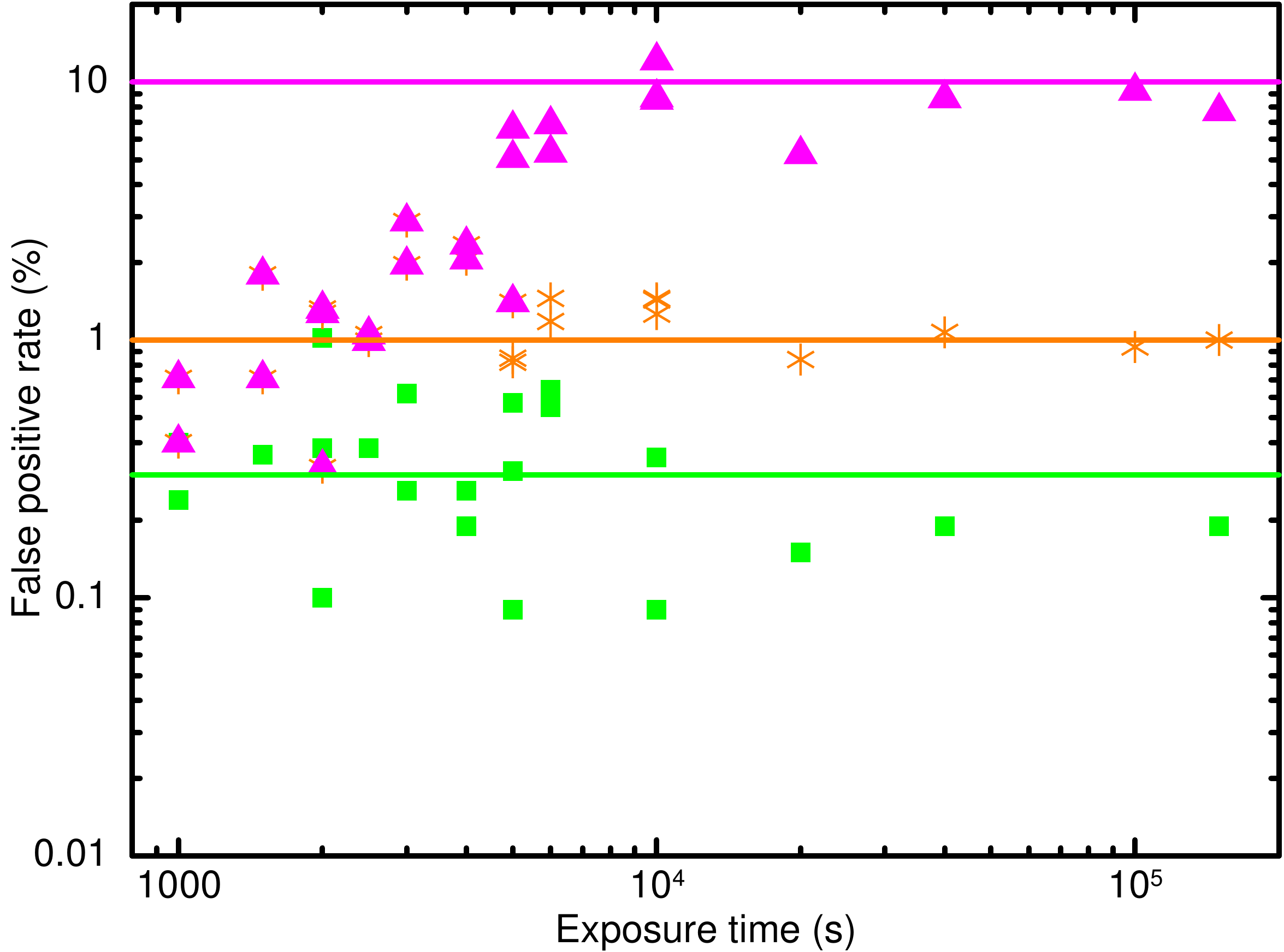}
\caption{The false positive rate measured from the various simulation runs, as a function of exposure time.
\emph{Green:} \good\ sources. \emph{Orange:} \good\ and \reasonable\ sources. \emph{Magenta} all sources.
The horizontal lines represent the 0.3\%, 1\%\ and 10\%\ levels.
}
\label{fig:falsep}
\end{center}
\end{figure}

The quality flags described in Section~\ref{sec:check} were calibrated 
such that the false positive rate in the catalog was 0.3\%, 1\%\ or 
10\%\ when \good, \good\ and \reasonable, or \good, \reasonable\ and 
\poor\ sources are included respectively. To calibrate these levels we 
again used simulations. Initially we performed a series of simulations 
of fixed exposure times (1,2,5,10,20,40 and 150 ks). We ran the catalog source detection 
software on each simulated image, and compared the list of detected 
sources with those simulated to determine the rate of false positives and therefore
set the likelihood thresholds corresponding to each quality flag. 
The false positive rate proved to be a function of exposure time, and we defined the quality flags accordingly.
To test these flag definitions over a range of exposures and background levels more representative
of the catalog than the discrete exposures use above, we ran a further 20,000 simulations, 
drawing the exposure time and background level at random from  the distribution of these values in the catalog datasets.
We found it necessary to reclassify some sources as \emph{Bad} based on their
positional errors. We also found that at exposures shorter than \til4 ks, the false positive rate never 
rose above \til2\%, we therefore added a caveat that, for images shorter than 4 ks, the flag could
only be \good\ or \reasonable. We ran a further 20,000 simulations to confirm that the results were stable.
The formal definitions of the flags are given in Table~\ref{tab:flagdef};  the false positive rate as a function
of exposure time and quality flag is shown in Fig.~\ref{fig:falsep}.

\begin{deluxetable*}{cl}
\tablecaption{Definitions of the detection flags}
\tablehead{
\colhead{Name}      &       \colhead{Definition} \\
} 
\startdata
Good  (=0)     & $L>18.52 E^{-0.051}$ \\
Reasonable (=1)  & $L\le18.52 E^{-0.051}$ ($E<4$ ks) \\
            & $L > 36.32 E^{-0.15}$   ( 4 ks $<E<40$ ks)\\
            & $L > 9.73 E^{-0.024}$   ($E\ge40$ ks)\\
Poor  (=2)      & $L > 86.55 E^{-0.29}$ (4 ks $<E<26$ ks) \\
            & $L > 3.47 E^{0.027}$   ($E\ge26$ ks)\\
Bad$^1$     & $L < L_{poor}$ or any position err ($\pm$RA,Dec) $>25\arcsec$ \\
\hline \\
Value=8    & As \good\ but in a region marked as containing artifacts. \\
Value=9    & As \reasonable\ but in a region marked as containing artifacts. \\
Value=10    & As \poor\ but in a region marked as containing artifacts. \\
Value=16    & As \good\ but in a region marked as containing diffuse emission. \\
Value=17    & As \reasonable\ but in a region marked as diffuse emission. \\
Value=18    & As \poor\ but in a region marked as diffuse emission. \\
\enddata
\tablecomments{$L$ is the source likelihood value, and $E$ the exposure time in seconds. \newline $^1$ \emph{Bad} detections
are not included in the catalog.}
\label{tab:flagdef}
\end{deluxetable*}

\begin{figure}
\begin{center}
\includegraphics[height=8.1cm,angle=-90]{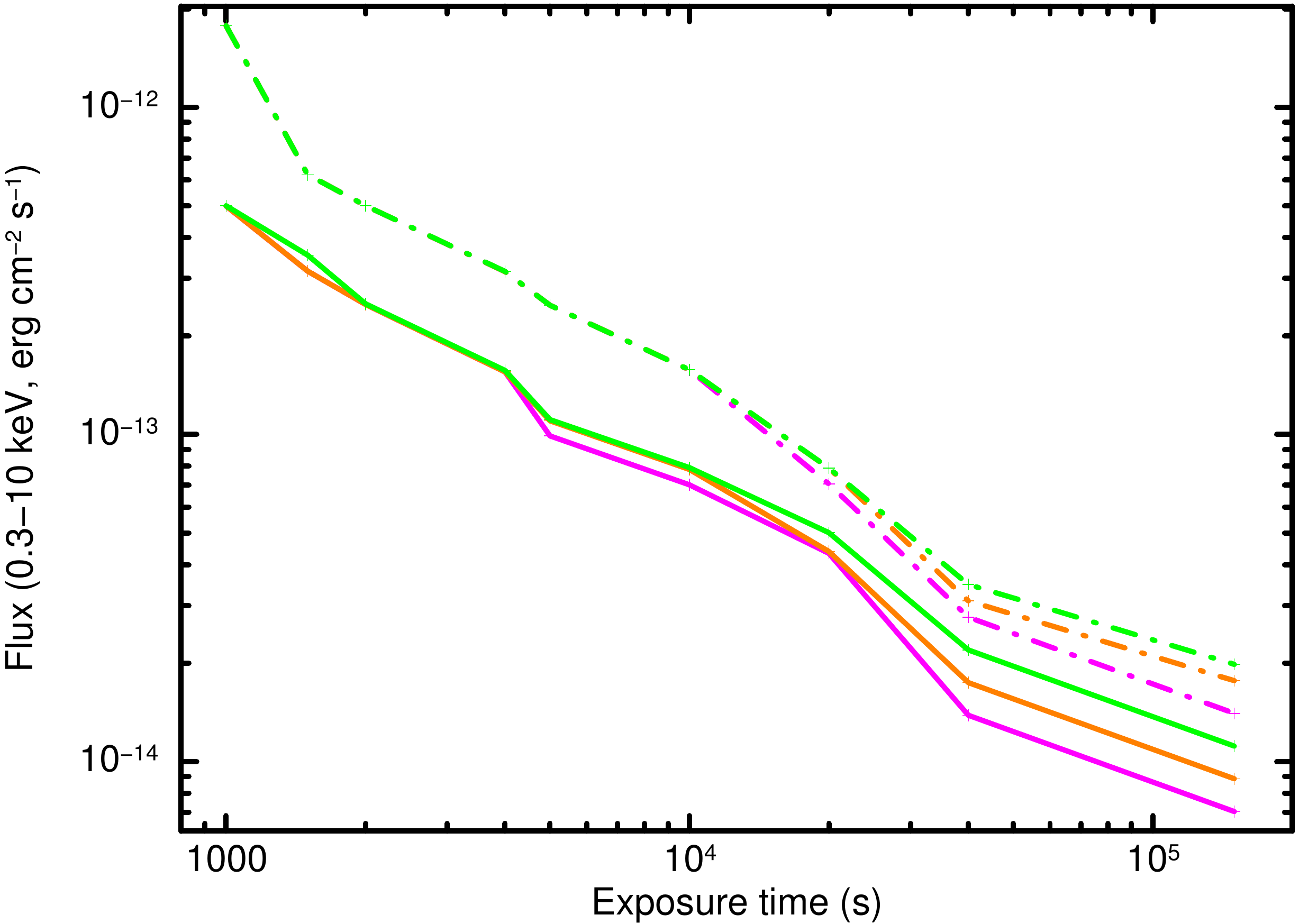}
\caption{The completeness of our detection method as a function of exposure time and quality flag.
The solid line is the 50\%\ complete level, the dot-dashed line the 90\%\ complete level.
\emph{Green:} \good\ sources. \emph{Orange:} \good\ and \reasonable\ sources. \emph{Magenta} all sources.
}
\label{fig:complete}
\end{center}
\end{figure}

\label{sec:complete}

We used the results of the simulations above to measure the fraction of simulated sources detected
as a function of 0.3--10 keV source flux, exposure time and quality flag. Fig.~\ref{fig:complete} shows
the result. The median exposure time of the observations in the catalog is 1.5 ks, at which 
our procedure is 50\%\ complete at 3\tim{-13} erg \cms\ s$^{-1}$, and 90\%\ complete
at 7\tim{-13} erg \cms\ s$^{-1}$. For the stacked images, the median exposure time
is 6 ks, at which our catalog is 50\%\ complete at 1\tim{-13} erg \cms\ s$^{-1}$, and
90\%\ complete at 2\tim{-13} erg \cms\ s$^{-1}$.

\section{Results and discussion}
\label{sec:discuss}

\begin{figure}
\begin{center}
\includegraphics[width=8.1cm]{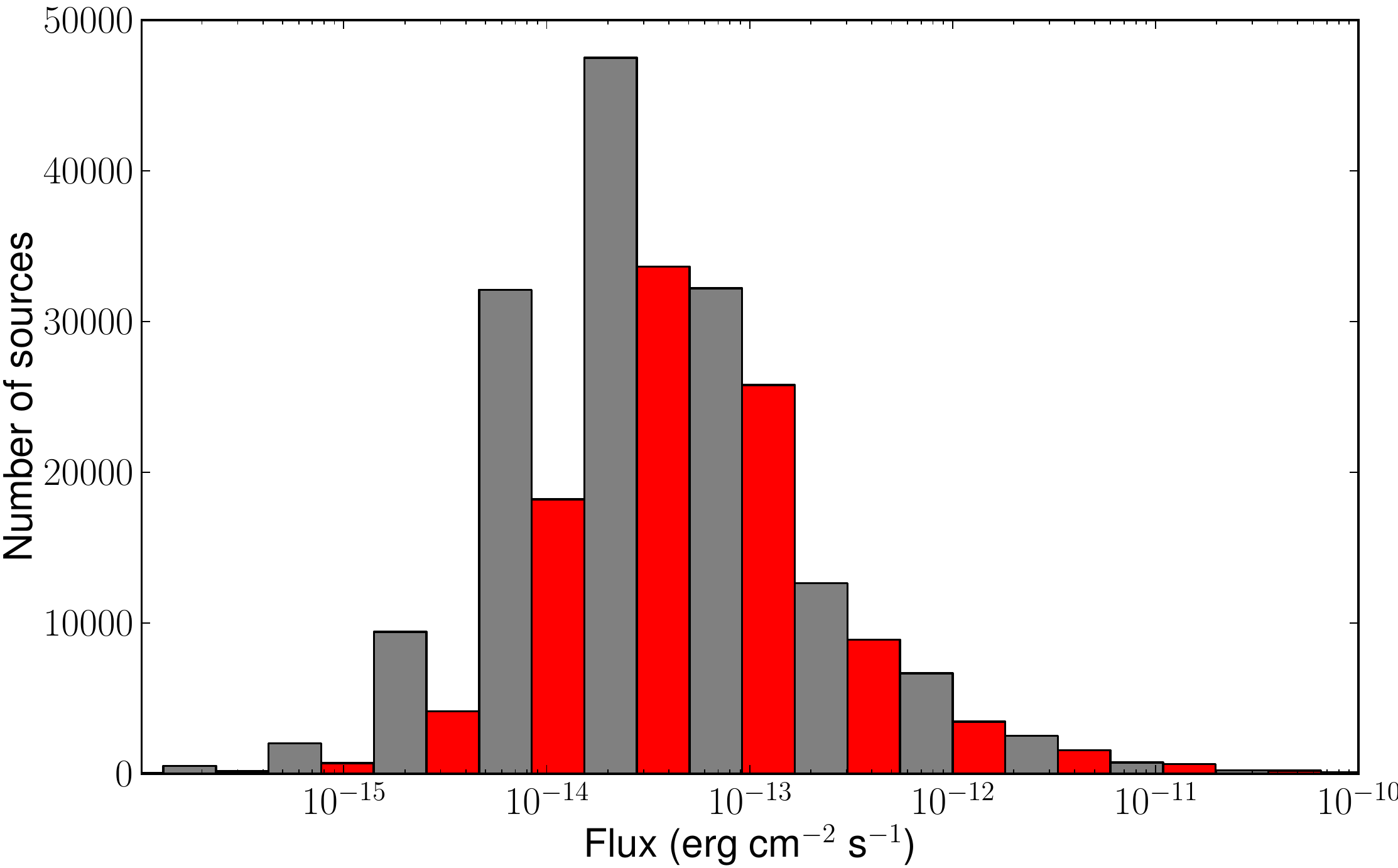}
\caption{The distribution of the 0.3--10 keV mean observed flux (derived assuming a power-law spectrum) for
the sources in our catalog. The gray bins are for all sources, the black
bins (red in the electronic version) for those in the `clean' subsample.
The flux shown is taken from the spectral fit, where available; otherwise it comes from the hardness ratio interpolation,
and if this is not available then from the fixed spectrum (see Section~\ref{sec:spec} for details).
}
\label{fig:flux}
\end{center}
\end{figure}

Our catalog contains 151,524 unique sources of which 98,762 
are in our highest quality `clean' subsample (\good\ and \reasonable\ 
sources only, excluding those in fields containing diffuse emission). 
Table~\ref{tab:flagstats} shows the breakdown of the sources according 
to the detection and field flags. The distribution of fluxes in the clean and total
samples is shown in Fig.~\ref{fig:flux}. Due to the effects 
of (in)completeness (Section~\ref{sec:falsep}) and the presence of the 
observation target object in our catalog, a $\log N-\log S$ calculation 
cannot be deduced directly from this figure -- see \cite{Mateos08} for a 
detailed discussion of the issues involved.

\begin{figure}
\begin{center}
\includegraphics[width=8.1cm]{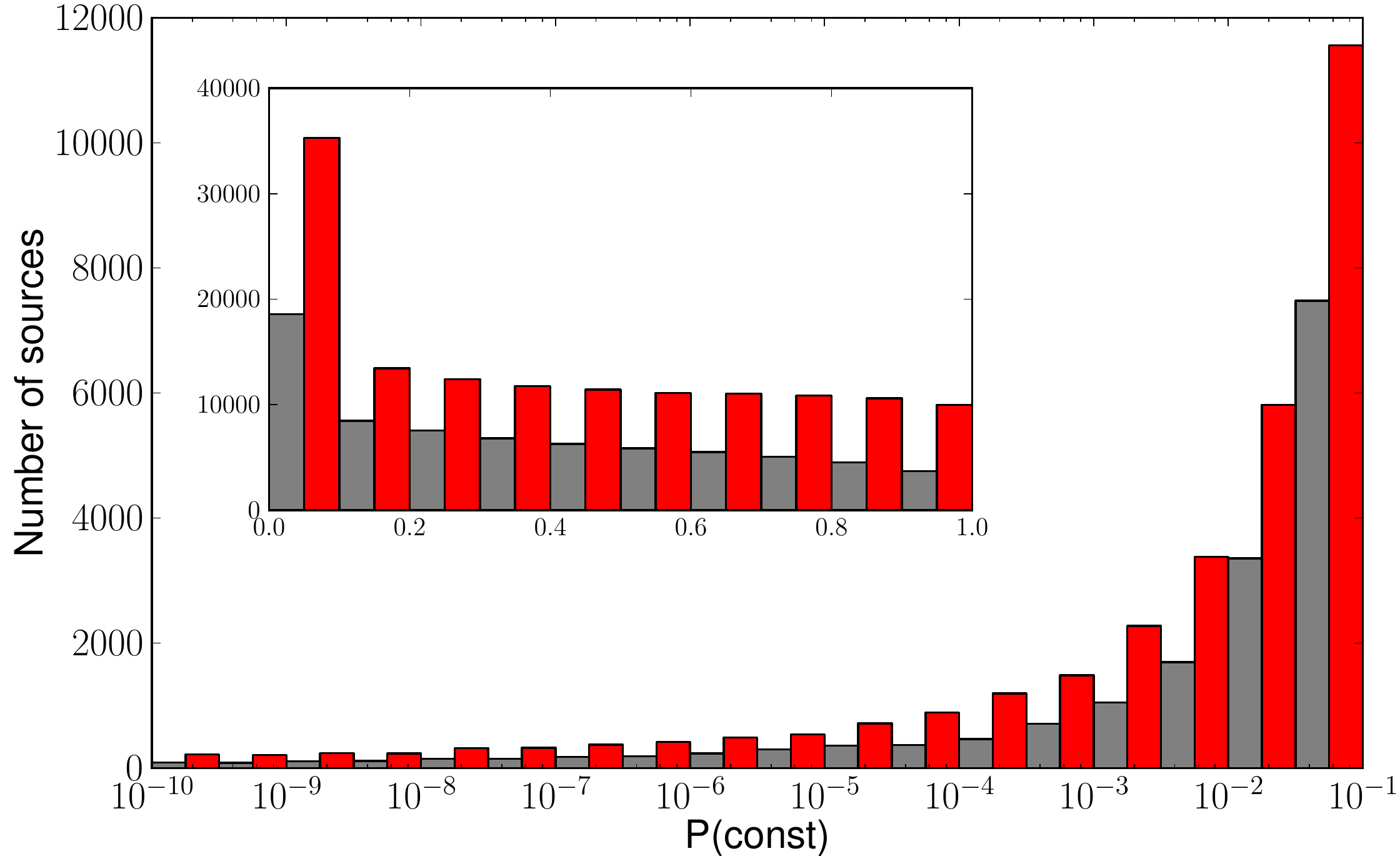}
\caption{The distribution of the \chisq-derived probability that a source is constant for the 1SXPS catalog sources,
excluding GRB afterglows. The gray data are for inter-snapshot
variability, the black bins (red in the electronic version) for inter-observation.
The inset shows the entire probability range, over which a population of constant sources would show equal numbers
of objects in each bin: the sharp spike at $P<0.1$ indicates a population of variable sources; the main plot
shows a magnified view (with a logarithmic probability axis) of this region.
}
\label{fig:varres}
\end{center}
\end{figure}

\begin{figure}
\begin{center}
\includegraphics[height=8.1cm,angle=-90]{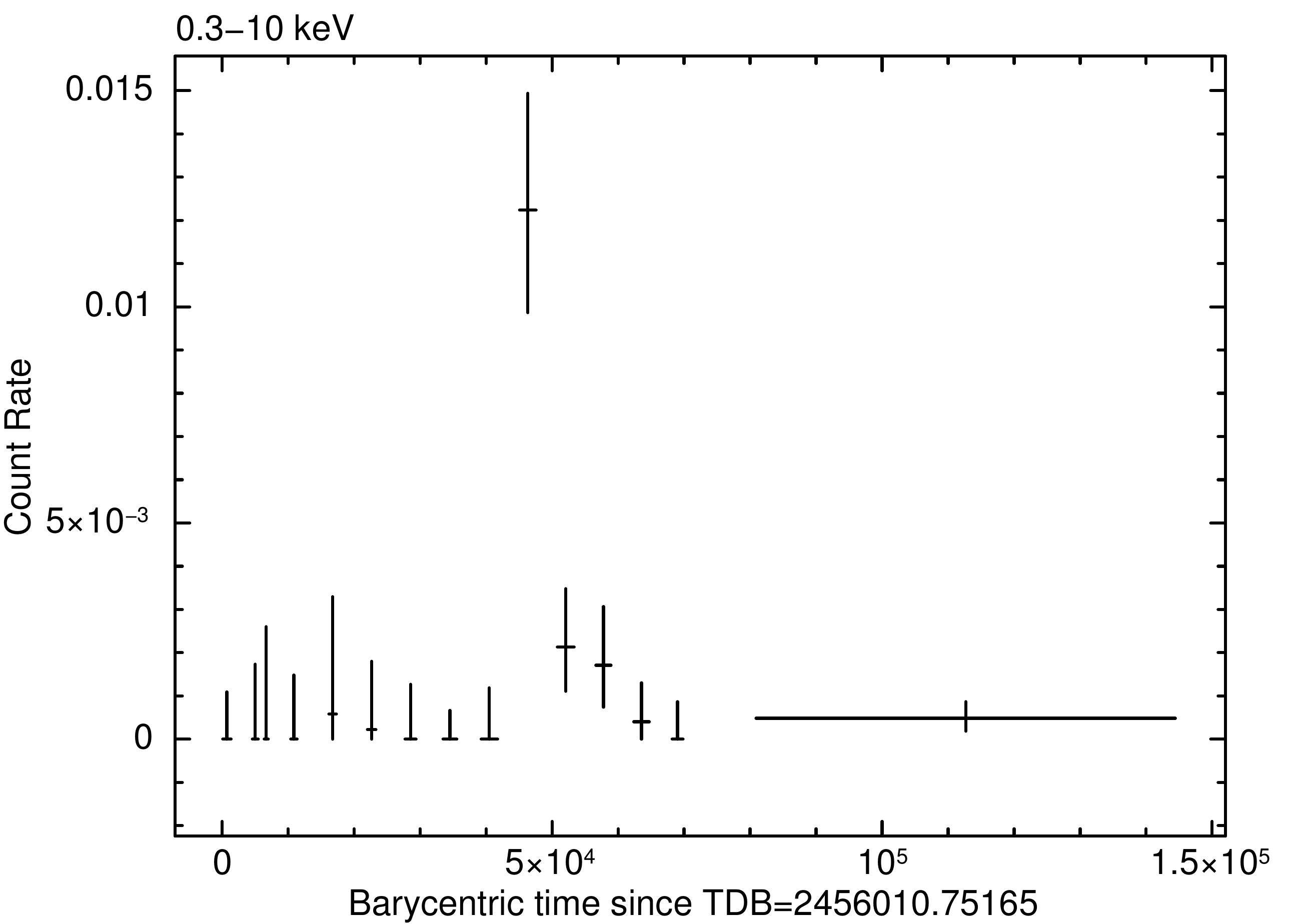}
\caption{The total-band light curve of 1SXPS J192427.2+240925 with one bin per snapshot. 
This is a short-lived transient, newly discovered in the 1SXPS catalog.}
\label{fig:egvar}
\end{center}
\end{figure}

Due to the observing strategy of \swift, our catalog gives a unique 
insight into variability on multiple timescales. Excluding GRBs, 28,906 
sources are found to be variable at the 3-$\sigma$ level in at least one 
band or binning method. Fig.~\ref{fig:varres} shows the distribution of 
the \chisq\ variability probability (Section~\ref{sec:temporal}) for the 
total-band light curves, GRBs have been excluded from this plot. A clear 
excess above the expected uniform distribution is seen at low 
probability of being constant, indicating a population of variable 
sources. Fig.~\ref{fig:egvar} shows an example light curve of one of 
these sources, 1SXPS J192427.2+240925, which appears to be short-lived 
transient that was only visible for three snapshots. This object was 
found by searching for sources in the clean catalog sample that had a 
low probability of being constant and no counterpart found in the 
external catalog cross-correlation (apart from a USNO-B1 or 2MASS 
object). Further investigation revealed a single $K=16.06$ mag stellar 
object in the UKIDSS Galactic Plane Survey in the XRT error region. This 
object is not in the USNO-B1 catalog, which has a limiting sensitivity 
of $V\til21$ mag. It thus seems likely that this object is an M dwarf 
star within 1 kpc, and that the XRT detected a coronal flare from it 
which lasted a few hours. 

\begin{deluxetable}{cc}
\tablecaption{The number of 1SXPS sources by flag values.}
\tablehead{
 \colhead{Flag Value}&       \colhead{Num Sources} \\
} 
\startdata
\multicolumn{2}{c}{\bf Detection flags} \\
\multicolumn{2}{c}{\emph{In fields flagged as OK}} \\
 \good & 69,967 (61\%) \\
 \reasonable & 16,127 (14\%) \\
 \poor & 27,904 (24\%) \\
\\
\multicolumn{2}{c}{\emph{In fields containing artifacts}} \\
\good & 9,856 (42\%) \\
\reasonable & 2,812 (12\%) \\
\poor & 5,557 (23\%) \\
Other$^1$ & 5,433 (23\%) \\
\\
\multicolumn{2}{c}{\emph{In fields containing diffuse emission}} \\
\good & 1,422 (10\%) \\
\reasonable & 455 (3\%) \\
\poor & 986 (7\%) \\
Other$^1$ & 11,005 (79\%) \\
\\
\multicolumn{2}{c}{\emph{In all fields}} \\
\good & 81,245 (54\%) \\
\reasonable & 19,394 (13\%) \\
\poor & 34,447 (23\%)\\
Other$^1$ & 16,438 (11\%) \\
\\

\hline

\multicolumn{2}{c}{\bf Field flags} \\

OK & 113,998  (75\%) \\
Has artifacts & 23,658 (16\%) \\
Has diffuse emission & 13,868 (9\%) \\
\enddata
\tablecomments{$^1$ `Other' refers to sources which lie within a region marked by manual screening,
i.e.\ sources with detection flags of 8 or above. See Section~\ref{sec:visscreen}.}
\label{tab:flagstats}
\end{deluxetable}

\begin{figure}
\begin{center}
\includegraphics[width=8.1cm]{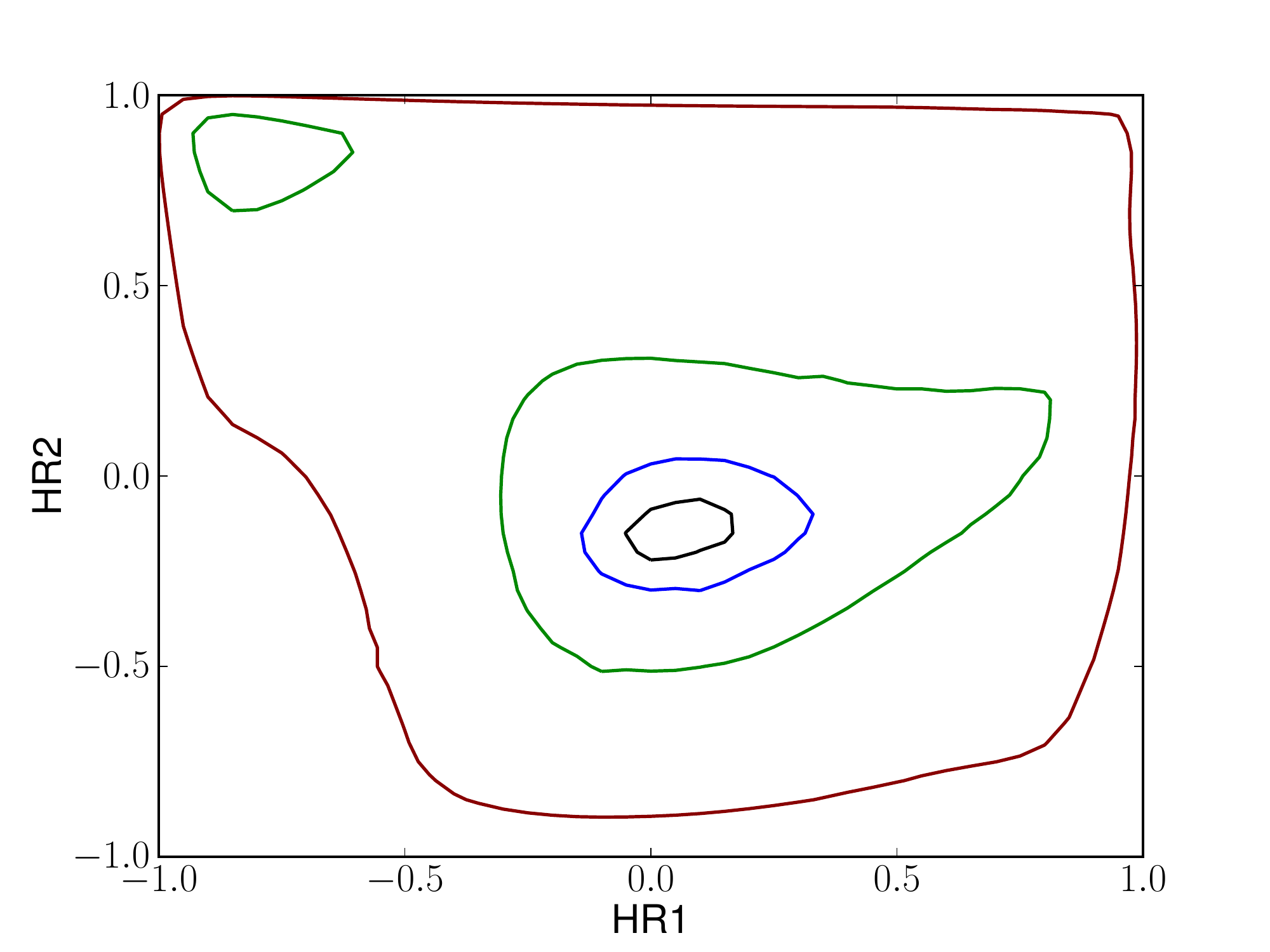}
\includegraphics[width=8.1cm]{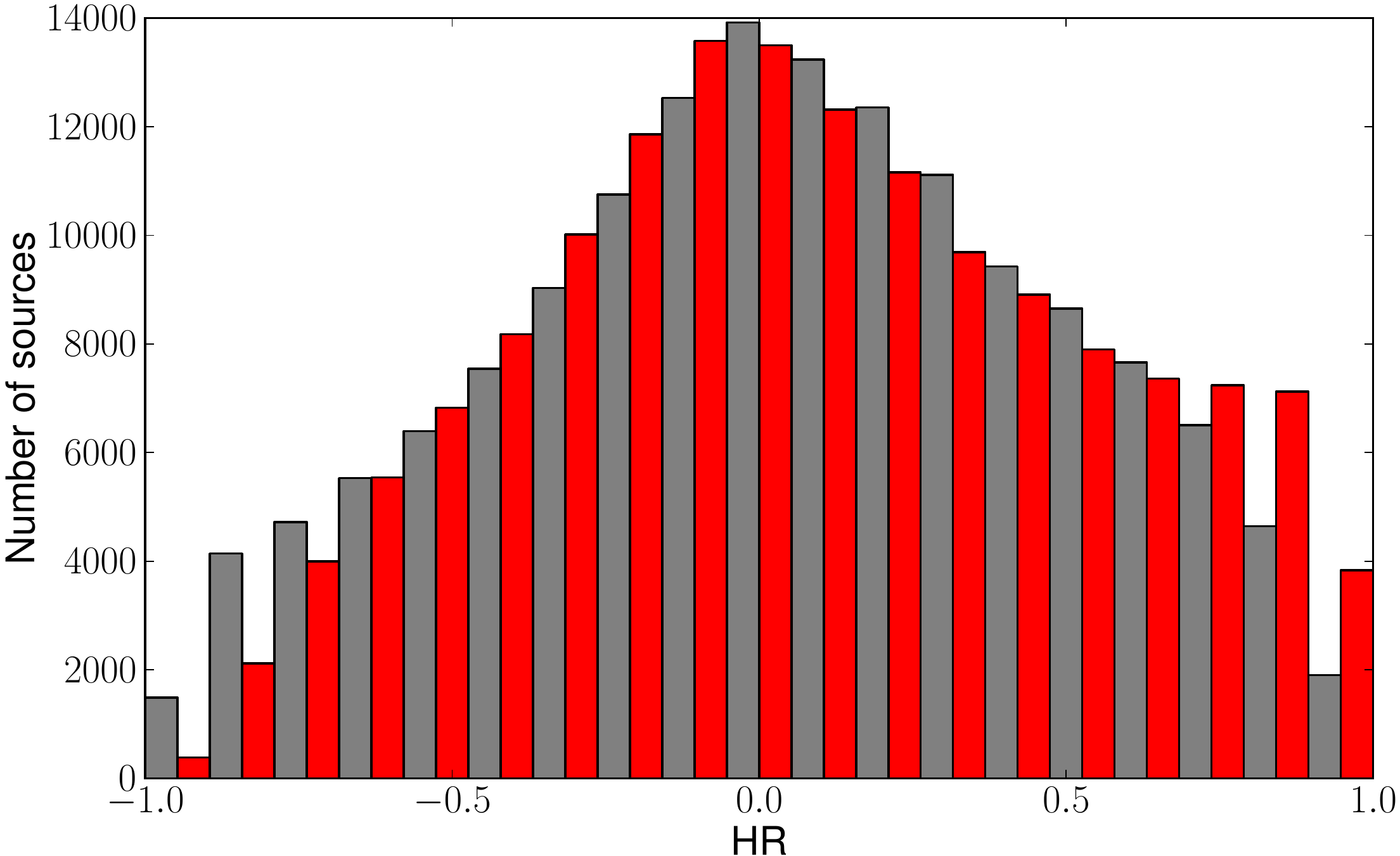}
\caption{\emph{Top:} Contour plot showing the (HR1,HR2) space occupied by the 1SXPS sources,
after smoothing over the error range of the individual sources. The 
contours show the areas 25\%,50\%,75\% and 90\% of the peak density.
\emph{Bottom:} The distribution of the individual HR values.
\emph{Gray:} HR1, \emph{Black:} HR2 (red in the electronic version).}
\label{fig:HRdensity}
\end{center}
\end{figure}

In Fig.~\ref{fig:HRinterp} we showed the area of (HR1,HR2) space permitted by simple spectral models
(a single absorber and emission component). Fig.~\ref{fig:HRdensity} shows the distribution of 
1SXPS sources in this space, revealing a significant number which do not lie within
the range permitted by these simple models. Indeed \til14,300 (9\%) of  all sources in the catalog
are not consistent with the single-component power-law or APEC models, at the 3-$\sigma$ level. Fig.~\ref{fig:HRdensity} 
also shows the distributions of the individual hardness ratios.

\subsection{Comparison with other catalogs}

The combination of sensitivity and sky coverage of this catalog 
means it occupies the area of parameter space 
between the deep-and-narrow surveys such as 3XMM-DR4, 1CSC \citep{iEvans10} and the 
\emph{Chandra} BMW catalog \citep{Romano08}; and the shallow-and-wide
surveys such as the \emph{Rosat} All-Sky Survey \citep{Voges99} and the 
XMM Slew Survey \citep{saxton08}. The number of sources in the 1SXPS catalog with no
counterpart in the set of catalogs shown in 
Table~\ref{tab:xcorr}\footnote{Excluding the 2MASS and USNO-B1 catalogs 
as the high spatial density of sources in these catalogs makes it hard to be certain
of association with the 1SXPS source.} is 68,638 (45\%) sources from the full catalog,
and 33,282 (34\%) sources in the clean sample. In part this is simply due to the limited overlap 
between surveys: 31\% of our fields have a 3XMM source 
within the field of view (including those undetected in our catalog), 
and 14\%\ have an SDSS quasar in the field, giving an idea of the size of 
the overlap.

The fields of the earlier XRT catalogs of \cite{puccetti11} and 
\cite{delia13} are included in the sample we have used. We found many 
sources not in those catalogs, partly because we included more data, but 
mainly because of the difference in strategy between the catalogs. 
\cite{puccetti11} focused on only stacked images of GRB fields 
(totaling 374 fields compared to our 7,343); \cite{delia13} used a much 
larger sample than \cite{puccetti11}, similar in size to ours (35,011 
observations compared to the 48,932 in our catalog\footnote{The 
difference arising partly because our sample extends ten months after 
the \cite{delia13} sample, and partly because we set a lower limit of 
100 s of PC mode data, where they use 500 s.}) however they used a 
higher SNR threshold than we did, and did not combine images thus 
limiting the sensitivity achieved. Our approach combines the advantages 
of both of these methods. Further, our improved detection system is 
significantly more sensitive than the {\sc ximage}-based approach 
employed in the earlier catalogs: for example simulations showed that in 
a 2 ks image for a source with a count-rate of 0.004 ct s$^{-1}$ 
(\til2\tim{-13} erg \cms\ s$^{-1}$, 0.3--10 keV) our system is 37\%\ 
complete, which is 1.5 times as complete as the {\sc ximage} system; the 
same is true for a source of count-rate 0.002 ct s$^{-1}$ 
(\til8\tim{-14} erg \cms\ s$^{-1}$, 0.3--10 keV) in a 5 ks image; the 
false positive rates in the two approaches were found to be similar. 
This combination of factors explains the number of sources present in 
our catalog that were not found in the earlier XRT catalogs.

%
%

The X-ray sky is highly variable, as evidenced by Fig.~\ref{fig:varres},
and to some extent all catalogs are biased in their contents towards sources in high states.
For example \cite{Starling11} used \swift\ to observe 94 unidentified X-ray sources from the 
\emph{XMM} slew survey with much greater sensitivity than that survey but only detected 30\%\ of
the \emph{XMM} objects. Nonetheless, this catalog, with its census of variability and useful combination of
moderate exposure and moderate sensitivity, will serve as a useful baseline for future missions
such as \emph{eRosita} and provides valuable information on the nature of variable sources which will
be part of the unresolved background for missions like \emph{LOFT}.

\section{Acknowledgements} 
PAE, JPO, APB, KLP, CP and CJM acknowledge support from the UK Space 
Agency. DNB and JAK acknowledge support from NASA contract NAS5-00136. 
GT acknowledges support from ASI-INAF grant I/004/11/0. We thank Simon 
Vaughan for helpful discussions during the preparation of this paper and 
Paul O'Brien for feedback on the manuscript. We also thank the anonymous referee
for insightful and helpful comments. This work made use of data 
supplied by the UK Swift Science Data Center at the University of 
Leicester. This research has made use of the XRT Data Analysis Software 
(XRTDAS) developed under the responsibility of the ASI Science Data 
Center (ASDC), Italy. This research has made use of the SIMBAD database,
operated at CDS, Strasbourg, France

\appendix
\section{Modifications to the PSF profile}
\label{sec:app}

\begin{deluxetable}{cc}
\tablecaption{The parameters for the PSF spoke model used in our background mapping tool}
\tablehead{
\colhead{Parameter}      & \colhead{Value} \\
}
\startdata
$u$ & 0.0574$^1$ \\
$v$ & 0.1512$^1$ \\
$R_{\rm min}$ & 5.6\arcsec \\
$R_{\rm pk}$ & 42.4\arcsec \\
$R_{\rm max}$ & 238\arcsec \\
$N_{\rm pk}$ & 0.21\\
\enddata
\tablecomments{$^1$ $u$ and $v$ are in units of half a phase, i.e. 15\deg.}
\label{tab:psfspoke}
\end{deluxetable}

The standard PSF of the \swift-XRT was calibrated by \cite{Moretti07} and is modelled as
a radially-symmetric King function:

\begin{equation}
P(R) \propto \left[1+ \left(\frac{R}{R_C}\right)^2\right]^{-\beta}
\end{equation}

\begin{figure}
\begin{center}
\hbox{
\includegraphics[width=12cm]{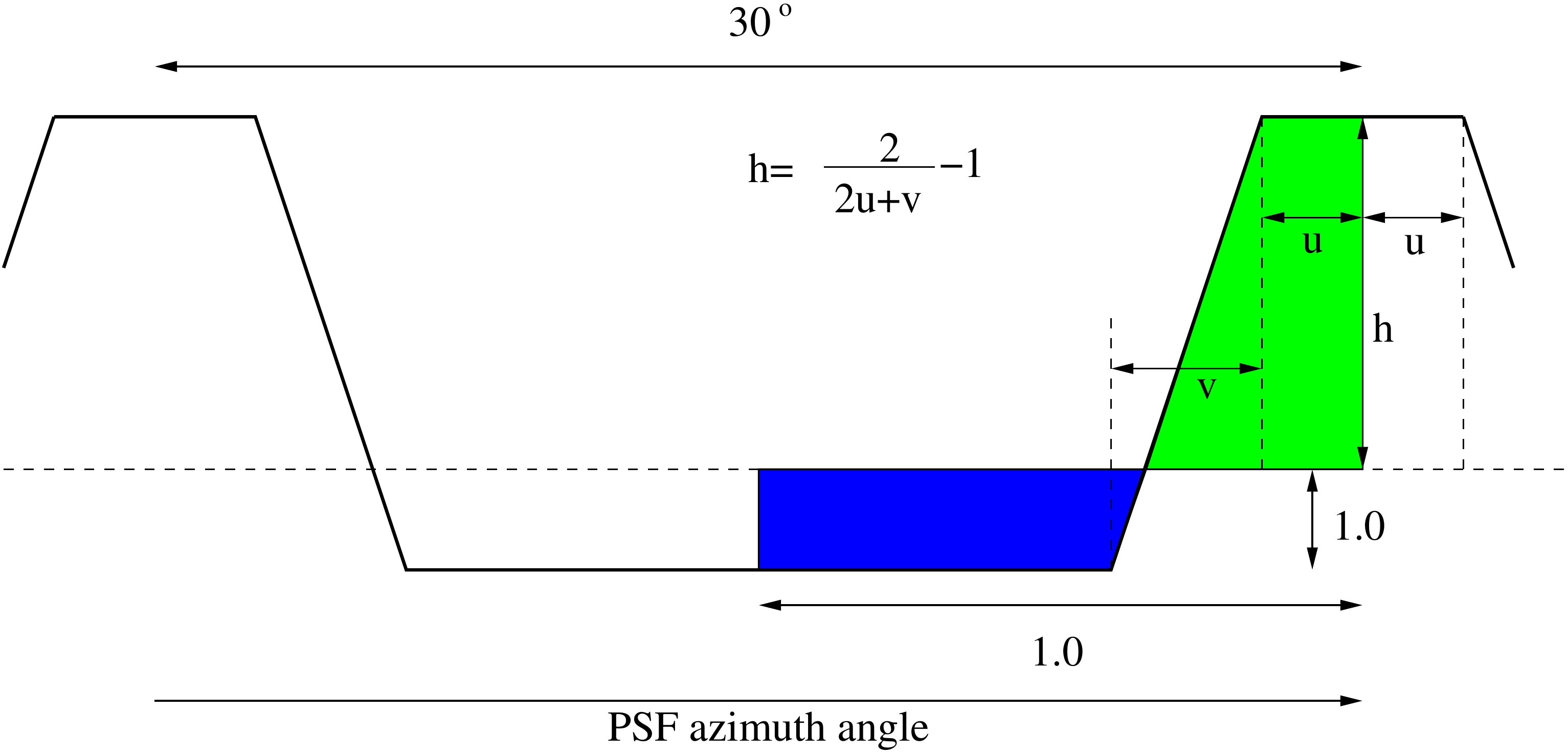}
\hspace{0.5cm}
\includegraphics[width=4cm]{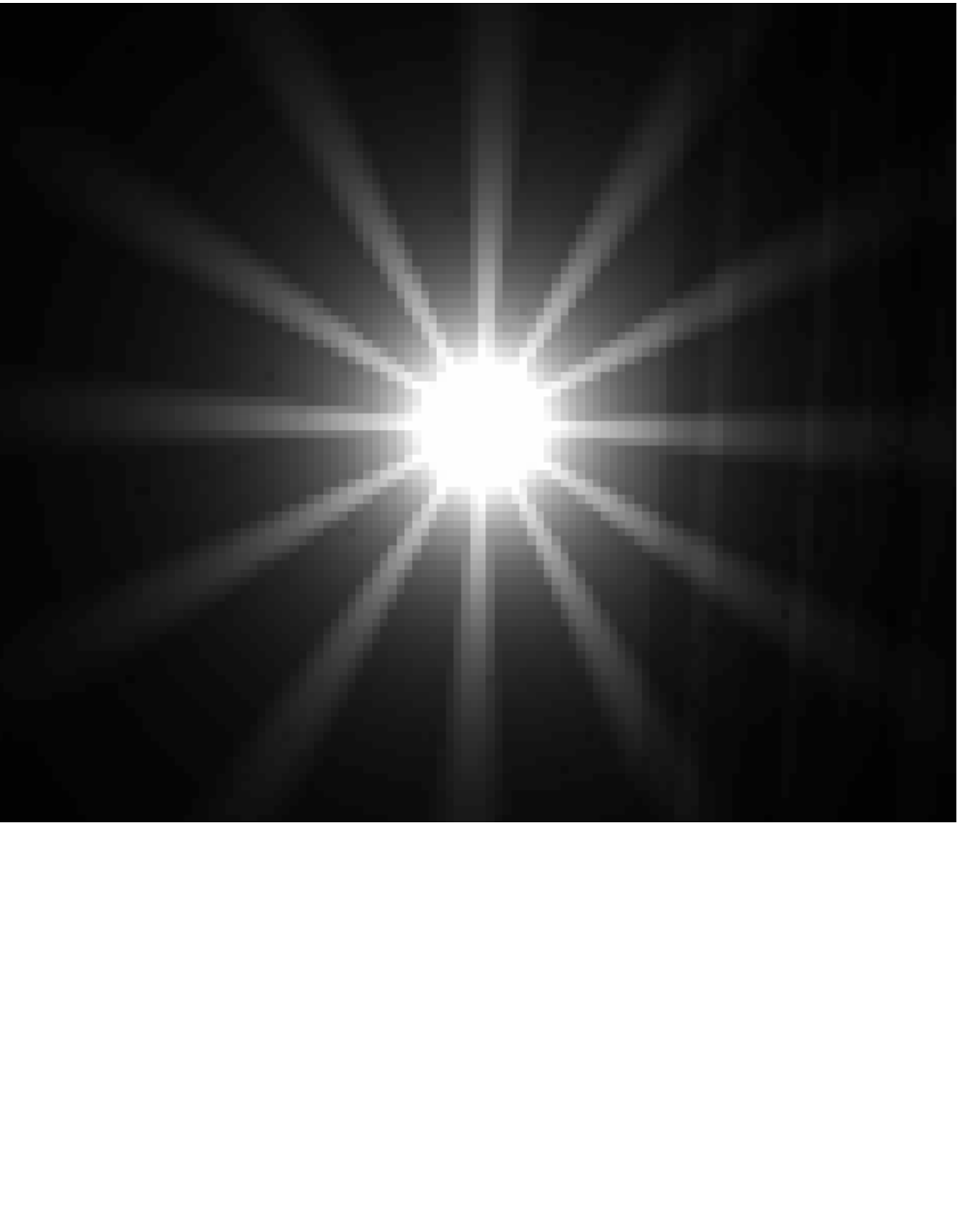}
}
\caption{\emph{Left:} the model for the PSF spokes, adapted from \cite{Read11}. The dashed horizontal line indicates the
level of the PSF without modulation by the spokes. The function is
constructed such that the dark and light gray 
areas have equal area (these are blue and green in the electronic
version). The model is normalized such that the horizontal measurements are in units of 
half the inter-spoke distance, i.e.\ 15\deg. and the vertical measurements are in units of the maximum reduction in PSF brightness. The figure is not to scale.
\emph{Right:} an example PSF model including the spokes. The intensity is logarithmically scaled. The non-radial structure is caused by the exposure map.}
\label{fig:psfmod}
\end{center}
\end{figure}

\noindent The real PSF shows deviations from this profile due to the presence of `spokes' caused by
the shadowing of light by the mirror support structure.
\cite{Read11} performed a comprehensive analysis of this effect for \emph{XMM}
and found that modulating the azimuthal variation of the PSF by a trapezoidal function, shown in Fig.~\ref{fig:psfmod},
gave a good representation of the PSF spokes. We applied this model to the XRT, first modifying it 
to account for the smaller number of spokes in XRT data (12, compared to 16 for \emph{XMM}); and then determined the function
parameters by fitting the model to an XRT dataset. If the model depicted in Fig.~\ref{fig:psfmod} is $f(\theta)$,
then the PSF is given by 

\begin{equation}
P(R,\theta) = P(R) \left[1 + N(R) f(\theta)\right]
\end{equation}

\noindent where $N(R)$ reflects the fact that the strength of the spoking effect is a function of radius within the PSF.
This is a simple function with four parameters: $N_{\rm pk}, R_{\rm min}, R_{\rm pk}$ and $R_{\rm max}$. 
$N(R)$ is given thus:
\begin{eqnarray}
N(R)  &=0 & (R<R_{\rm min}\ {\rm or}\ R>R_{\rm max}) \nonumber \\
N(R)  &=\left(\frac{N_{\rm pk}}{R_{\rm pk}-R_{\rm min}}\right) \left(R-R_{\rm min}\right) & (R_{\rm min}<R<R_{\rm pk}) \nonumber \\
N(R)  & = N_{\rm pk} - \left(\frac{N_{\rm pk}}{R_{\rm max}-R_{\rm pk}}\right) \left(R-R_{\rm pk}\right)  & (R_{\rm pk}\le R\le R_{\rm max})  \\
\end{eqnarray}

At $R<R_{\rm min}$ or $R>R_{\rm max}$
$N(R)=0$, at $R_{\rm min}<R<R_{\rm pk}$ $N(R)$ increases linearly to $N_{\rm pk}$ and then it decreases
linearly again to 0 at $R_{\rm max}$.

A non-piled up point source would ideally be used to fit the PSF spoke parameters however this proved impossible.
Each snapshot of \swift\ data has a slightly different pointing position and roll angle so to model the PSF spokes we had
to use only a single snapshot of data. Pile up becomes an issue at around 0.6 ct s$^{-1}$
so single-snapshot images of non piled-up sources did not contain enough counts
for us to perform a reliable fit to the relatively weak 
PSF spoke effect. We therefore used a brighter but piled up source, accepting that this will give us a model to the PSF spokes
which is probably imperfect for non-piled-up source, but better than no model at all. The parameters of this fit
are given in Table~\ref{tab:psfspoke}.

As well as adding in the PSF spokes it was sometimes necessary to incorporate out-of-time events into the background
map when modeling sources. Out-of-time events are events detected while the CCD is being 
read out, spreading the $y$-position of those events along the entire column. Since the deadtime
in PC mode is equal to 0.004 times the exposure time, the count-rate of out-of-time events in a given CCD column
is simply 0.004 times the number of in-time events in that column. We estimate the latter value by reading the number
of events in a 41 pixel high region centered on the source and then multiply this by 0.004 and divide it by 600 (the number of rows
on the CCD). We then add the resultant value to the background map for every pixel on that row. We perform this for an 11 
pixel wide region centered on the source. This is only done for sources brighter than 3 ct s$^{-1}$ since below this level OOT
events are insignificant compared to the background.

\bibliographystyle{apj}
\bibliography{phil}

\label{lastpage}

\end{document}